\documentclass[pre,twocolumn,showpacs,preprintnumbers,amsmath,amssymb,floatfix]{revtex4-1}

\usepackage{epsfig,graphicx}
\usepackage{psfrag}
\usepackage[normalem]{ulem}
\usepackage{color}
\usepackage{bm}
\def\s#1{_{\rm #1} }
\def\sp#1{^{\rm #1} }

\def\n{{\hat{\uline{n} }}}
\def \be{\begin{equation}}
\def \bea{\begin{eqnarray}}
\def \ee{\end{equation}}
\def \eea{\end{eqnarray}}

\def\sp{s_{\phi}}
\def\cp{c_{\phi}}

\def\s2p{s_{2\phi}}
\def\c2p{c_{2\phi}}
\def\s2ps{s_{2\phi^{*}}}
\def\c2ps{c_{2\phi^{*}}}
\def\st{s_{\theta}}
\def\ct{c_{\theta}}
\def\Kt{\tilde{K}}
\def\zt{\tilde{z}}

\def\bea{\begin{eqnarray}}
\def\eea{\end{eqnarray}}
\def \be{\begin{equation}}
\def \ee{\end{equation}}

\def \et{\textit{et. al.} }

\begin{document}
\title{Modelling the helical-flexo-electro-optic effect}
\author{D. R. Corbett}
\email{daniel.corbett@gmail.com}
\address{Department of Engineering Science, University of Oxford, Parks Road, Oxford OX1 3PJ, U.K.}
\author{S. J. Elston}
\address{Department of Engineering Science, University of Oxford, Parks Road, Oxford OX1 3PJ, U.K.}
\date{\today}
\begin{abstract}
The helical-flexo-electro-optic effect shows interesting in-plane electro-optic switching behaviour due to flexoelectric coupling with applied electric fields.  Previous understanding of the behaviour has been generally based on an analytic approach which makes certain assumptions about the uniformity of the helical structure and the induced tilt angle under field application.  Here we remove these assumptions and develop a perturbative approximation to describe the structure in more detail.  We also use a numerical method to investigate the behaviour in regimes where the perturbative approach is inappropriate.  The impact of variation in elastic constants and dielectric anisotropy is investigated.  We find that dielectric behaviour in particular can lead to substantial differences between the tilt angle obtained here and those obtained using previous analytic models.
 \end{abstract}
\maketitle
\section{Introduction}\label{sec:one}

Flexoelectricity in liquid crystals has been the subject of interest since Meyer first discussed the effect in 1969~\cite{Meyer:69}.  It leads to a direct coupling between splay/bend distortions and polarisation in nematic materials.  The flexoelectric polarisation is described by:
\begin{equation}
\uline{P}_{flexo}=e_{1}\n(\nabla\cdot\n)+e_{3}(\nabla\times\n)\times\n,\label{eq:flexpol}
\end{equation}
where $\n$ is the (unit magnitude) nematic director, $e_{1}$ is the splay flexoelectric coefficient and $e_{3}$ is the bend coefficient.  No polarisation results from a twist distortion~\cite{Rudquist:97}, hence the absence of a coefficient $e_{2}$.  Interaction between this polarisation and electric fields results in a direct link between fields and director curvature.  This is somewhat different to the coupling between electric fields and the dielectric anisotropy of liquid crystals, which promotes director reorientation rather than director curvature itself.  Director curvature might result from this latter effect due to competition with elasticity in the liquid crystal, but it is not a direct coupling between electric field and director curvature.

\begin{figure}
\includegraphics[width=0.4\textwidth]{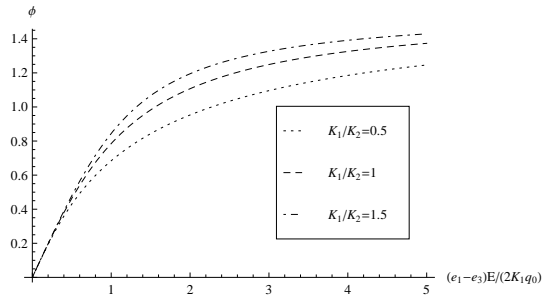}
\caption{The variation of the tilt angle $\phi$ with~$(e_{1}-~e_{3})E/(2K_{1}q_{0})$.  All three lines have $K_{1}=K_{3}$.  The plot for $K_{1}=K_{2}$ shows the result from equation~\ref{eq:meyer} while the other two plots are from equation~\ref{eq:lee}. }
\label{fig:compare}
\end{figure}

In the original work on the flexoelectric effect it was suggested that a splay-bend pattern could lead to a net polarisation, and that conversely applying an electric field should generate a splay-bend distortion.  In 1987 such a splay-bend pattern was linked to a distortion in a chiral nematic material, leading to the discovery of the helical-flexo-electro-optic effect~\cite{Patel:87}, which occurs when an electric field is applied perpendicular to the helical axis in a cholesteric.  For example, in a liquid crystal device with a chiral nematic aligned with the helix axis parallel to the surfaces (the uniformly lying helix, or ULH structure) an electric field applied between the device surfaces satisfies this condition.  Patel and Meyer~\cite{Patel:87} showed in this case that the application of an electric field leads to a rotation of the directors by an angle $\phi$ in the plane perpendicular to the applied field, given by:
\begin{equation}
\tan \phi=\frac{(e_{1}-e_{3})E}{2Kq_{0}},\label{eq:meyer}
\end{equation}
\begin{figure}
\includegraphics[width=0.45\textwidth]{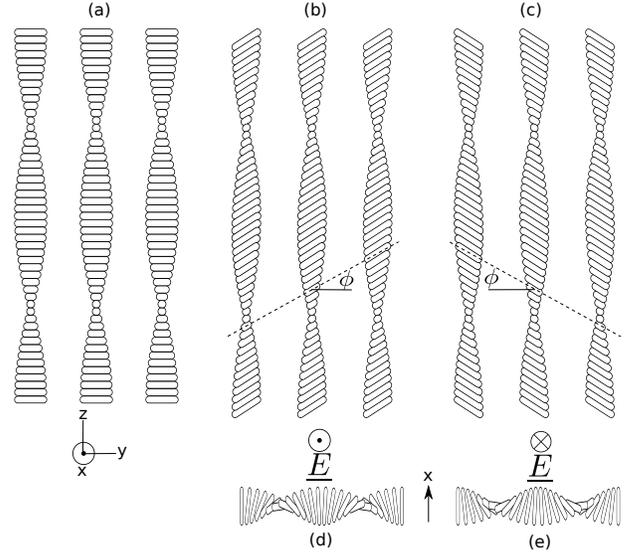}
\caption{The helical-flexo-electro-optic effect.  (a) Initially we have a helical structure twisting in a right handed sense with the helical axis along the $z$-direction (b) applying an electric field out of the page ($x$-axis) causes the directors to rotate about the electric field in a right handed sense by an angle $\phi$. (c) Reversing the direction of the field reverses the direction of the rotation. (d) Investigating the director structure along the dotted line in (b) we see a characteristic splay-bend pattern which results in a flexoelectric polarisation. (e) Similarly following the dotted line in (c) we see the a reversed splay-bend pattern.}
\label{fig:one}
\end{figure}where $e_{1}$ and $e_{3}$ are the splay and bend flexoelectric coefficients, $K$ is an average Frank elastic constant, $E$ is the magnitude of the applied electric field and $q_{0}$ is the magnitude of the wave-vector of the undistorted helix.  The behaviour is illustrated in figure~\ref{fig:one}(b), an electric field is applied perpendicular to the helix of a cholesteric, resulting in a director reorientation.  Tracing the director along the dotted line inclined at angle $\phi$ gives the splay-bend pattern shown in figure~\ref{fig:one}(d).  Reversing the direction of the electric field reverses the direction of the tilt, see figure~\ref{fig:one}(c) and (e).  Patel and Meyer also provided a slightly more general result in the case where the splay ($K_{1}$) and bend ($K_{3}$) elastic constants are equal, but are in general different from the twist constant $K_{2}$.  In this case the constant $K$ appearing in equation~\ref{eq:meyer} is replaced by $K_{1}$.  One of the key assumptions underlying equation~\ref{eq:meyer} is that the helical pitch is allowed to vary freely in order to minimise the free energy per unit pitch of the system.  Flexoelectric interactions then lead to a decrease in the helical pitch with increasing electric field.  In a ULH device it is likely that the pitch is not free to vary, for example both Rudquist \et~\cite{Rudquist:98a} and Broughton \et~\cite{Coles:06} have used polymerisation to increase the stability of the ULH structure relative to the Grandjean texture.  The fixed pitch case has been considered by both Lee \et~\cite{Lee:90} and Rudquist \et~\cite{Rudquist:94}, who obtain:
\begin{equation}
\tan \phi = \frac{(e_{1}-e_{3})E}{2K_{2}q_{0}}-\frac{(K_{1}+K_{3}-2K_{2})}{2K_{2}}\sin\phi.\label{eq:lee}
\end{equation}
We shall see later in section~\ref{sec:foura} that the assumption of a constant angle $\phi$ upon which the equation relies is only valid if $K_{1}=K_{3}$.  Equation~\ref{eq:lee} gives a different variation for $\phi$ with $E$ when compared with equation~\ref{eq:meyer}.  Note however, for small angles both equations give a similar linear response with the electric field $\phi\approx(e_{1}-e_{3})E/(2K_{1}q_{0})$ (with $K_{1}=K_{3}$).  Figure~\ref{fig:compare} shows a plot of the tilt angle $\phi$ as a function of $(e_{1}-e_{3})E/(2K_{1}q_{0})$ for both equations~\ref{eq:meyer} and~\ref{eq:lee} for the case where $K_{1}=K_{3}$.  Provided $K_{1}>K_{2}$ (which is true for most liquid crystal materials) the tilt angle $\phi$ from equation~\ref{eq:lee} will be greater than that from equation~\ref{eq:meyer}.
There are three key assumptions underlying the derivation of equation~\ref{eq:lee}.  Firstly, it is assumed that the fixed pitch is the same as that implied by the intrinsic chirality of the liquid crystal mesogens, however this need not be so. For example, polymer stabilisation is quite likely to fix the value of the pitch to the value prevailing at the temperature of polymerisation; changes in temperature will alter the preferred intrinsic pitch, leading to a suboptimal fixed pitch.  Secondly it is assumed that the underlying helix remains undistorted.  The helix can in principle distort in two ways.  Regions of the helix can be compressed and other regions dilated so that the helix is no longer uniform, but the overall pitch remains unchanged.  Additionally, the overall pitch of the helix could change.  Helix distortion is highly likely if dielectric effects are important, but could also arise as a result of differences in the elastic constants.  Thirdly it has been assumed that the reorientation of the director about the electric field can be described by a single angle $\phi$, which is independent of position along the helix.  This is a somewhat severe constraint, at different points along the helix the reorienting torque is parallel to or perpendicular to the director.  It is not clear that the net reorientation should be the same in both places, and therefore reorientation which is spatially non-uniform may be possible. 

In this paper we remove these three key assumptions in order to model the helical-flexo-electro-optic effect in more detail.  We thus allow for the pitch of the helix to be different from the intrinsic pitch implied by the chirality of the liquid crystal mesogens.  Furthermore we allow the helix to distort.  As mentioned, the helix can distort both by local compressions and dilations, and by an overall change in the pitch.  In this work we consider the latter effect only for the special cases of the one elastic constant model in section~\ref{sec:three} and the two elastic constant model in section~\ref{sec:four}.  In these two special cases exact analytic solutions are possible.  For more general elastic constants we keep the overall pitch fixed.  This is principally to make the problem more tractable, but is potentially reasonable because in practical situations the overall pitch is likely to be pinned within the ULH structure, but local distortions can take place.  Finally, and perhaps most crucially, we do not constrain the tilt angle $\phi$ to be a constant along the helix, but rather we let it vary.  This paper is organised as follows, in section~\ref{sec:two} we introduce a free energy functional which includes Frank elastic, flexoelectric and dielectric  contributions.  The Euler-Lagrange (EL) equations which minimise this free energy functional are derived.  In section~\ref{sec:three} we solve the EL equations for the simplified one elastic constant case.  In section~\ref{sec:four} we generalise the result obtained by both Lee and Rudquist shown in equation~\ref{eq:lee} to include the effects of a difference between the intrinsic pitch $L_{0}$ and the fixed pitch $L$.  In section~\ref{sec:foura} we present a perturbative solution for the angle $\phi$ for the general case with three separate elastic constants and with dielectric effects.  In section~\ref{sec:five} we provide numerical results showing $\phi$ as a function of position along the helix for a variety of elastic and dielectric parameters.  Finally in~\ref{sec:six} we present conclusions.

\section{Modelling}\label{sec:two}

We assume that the undistorted director structure is a pure twist deformation with the helix axis along the $z$-axis and with a period $L$, which is in general different from the pitch $L_{0}=2\pi/q_{0}$ implied by the intrinsic chirality $q_{0}$.  To this undistorted structure we apply an electric field along the $x$-axis.  The free energy density is:
\begin{eqnarray}
f&=&\frac{K_{1}}{2}(\nabla\cdot\n)^{2}+\frac{K_{2}}{2}(\n\cdot\nabla\times\n+q_{0})^{2}+\frac{K_{3}}{2}(\n\times\nabla\times\n)^2\nonumber\\&&-\uline{E}\cdot\uline{P}_{flexo}-\frac{1}{2}\epsilon_{0}\Delta\epsilon(\n\cdot\uline{E})^{2},\label{eq:freenonscaled}
\end{eqnarray}
where $\uline{P}_{flexo}$ is the flexoelectric polarisation given in equation~\ref{eq:flexpol}, $\uline{E}=(E,0,0)$ is the electric field, $\epsilon_{0}$ is the permittivity of free space and $\Delta \epsilon$ is the dielectric anisotropy of the material.  The sign of the flexoelectric and dielectric additions is given by taking into account the work done by the external voltage source to maintain the electrode potentials.  We assume that the director varies only along the $z$-direction, thus $\nabla\equiv(0,0,\partial_{z})$, where $\partial_{z}$ indicates differentiation with respect to the coordinate $z$.  We shall focus only on bulk behaviour, thus terms within the free energy which can be transformed into surface contributions will be ignored.  Finally we shall assume the electric field is constant and will not account for the updates in it which occur as a result of changes in the director field.  Flexoelectric torques naturally divide into effects which depend on the sum of the flexoelectric coefficients $(e_{1}+e_{3})$ and their difference $(e_{1}-e_{3})$.  The sum terms couple to gradients in the electric field while the difference terms couple to gradients in the director, thus our assumption of a constant electric field implies we will only see flexoelectric terms that depend on the combination $(e_{1}-e_{3})$~\cite{petrov}.

The free energy per unit pitch functional is then conveniently re-written as:
\begin{eqnarray}
{\cal F}&=&\int_{0}^{L}\left\{K_{1}\left[\uline{S}-\frac{e_{1}-e_{3}}{2K_{1}}\uline{E}\right]^{2}+K_{3}\left[\uline{B}-\frac{e_{1}-e_{3}}{2K_{3}}\uline{E}\right]^{2}\right.\nonumber\\&&+K_{2}(T+q_{0})^{2}-\left.\epsilon_{0}\Delta\epsilon(\n\cdot\uline{E})^{2}\right\}\left(\frac{dz}{2L}\right),\nonumber\\
\label{eq:free}
\end{eqnarray}
\begin{figure}
\includegraphics[width=0.45\textwidth]{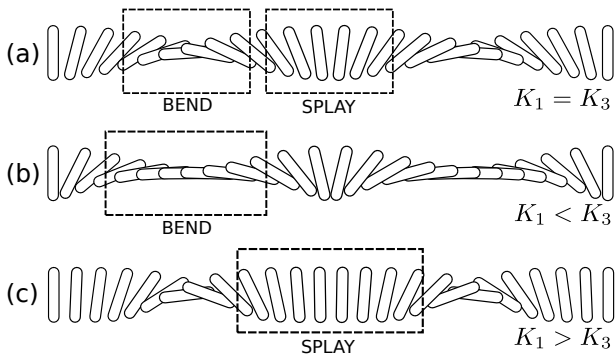}
\caption{The splay-bend pattern in for (a) $K_{1}=K_{3}$, regions of high splay and bend distortion are highlighted. (b) $K_{3}>K_{1}$, in this case the region of bend distortion is extended and (c) $K_{1}>K_{3}$ in this case the region of splay distortion is extended.}
\label{fig:onea}
\end{figure}where we have introduced the splay vector $\uline{S}=\n(\nabla\cdot\n)$, the bend vector $\uline{B}=(\nabla\times\n)\times\n$ and the twist pseudoscalar $T=\n\cdot\nabla\times\n$.  We note that the bend and splay vectors are always mutually orthogonal, i.e. $\uline{S}\cdot\uline{B}=0$.  We are now in a position to understand the effect differences between $K_{1}$ and $K_{3}$ will have on the equilibrium structure.  In figure~\ref{fig:onea}(a) we show the splay-bend pattern that arises when $K_{1}=K_{3}$.  As we can see the structure involves an oscillation between a bend and a splay distortion as one progresses along the helix.  Regions of bend and splay are highlighted.  In figure~\ref{fig:onea}(b) we show a schematic of the structure if $K_{3}>K_{1}$.  In this case we can see that the optimal induced bend $(e_{1}-e_{3})/(2K_{3})\underline{E}$ is lower than the optimal induced splay $(e_{1}-e_{3})/(2K_{1})\underline{E}$, and deviations from the optimal bend are more heavily penalised than deviations from the optimal induced splay (since $K_{3}>K_{1}$).  We therefore expect the overall magnitude of the induced bend to be reduced, and the spatial region over which bend is the predominant distortion to be extended.  Reversing the argument if $K_{1}>K_{3}$ we expect the magnitude of the induced splay to be reduced, and the spatial region over which splay is the predominant distortion to be extended, as shown in the figure~\ref{fig:onea}(c).  It is interesting to note that the effect of $K_{1}>K_{3}$ is similar to that of positive dielectric anisotropy - the director reorients towards the electric field. Similarly $K_{3}>K_{1}$ has an effect similar to a negative dielectric anisotropy, the director reorients such that it is perpendicular to the field.



\subsection{Sign Conventions and Scalings}

We have chosen to write the twisting contribution to the free energy density as $\frac{K_{2}}{2}(T+q_{0})^{2}$ rather than $\frac{K_{2}}{2}(T-q_{0})^{2}$.  The latter convention was used by Patel and Meyer~\cite{Patel:87}, we prefer the former since since a positive value of $q_{0}$ then corresponds to a right handed helix, this is the standard convention in the literature.  We have chosen to write the bend vector as $\uline{B}=(\nabla\times\n)\times\n$ rather than $\uline{B}=\n\times(\nabla\times\n)$.  This convention is that adopted by de Gennes and Prost~\cite{deGennesprostbook:93}.  With this choice the bend vector points inwards towards the centre of bend, rather than outwards away from the centre of bend.  With these two sign conventions we will see that the director in a right handed helix (+ve $q_{0}$) will experience a right handed reorienting torque about the electric field. 

In order to simplify the resulting EL equations we scale all lengths by the imposed pitch $L$ and the free energy density by $(K_{1}+K_{3})/(2L_{0}^{2})$.  These scalings, and several important dimensionless variables are listed here for convenience:
\begin{eqnarray}
\tilde{{\cal F}}=\frac{2L_{0}^{2}}{(K_{1}+K_{3})}{\cal F}&;\,&\chi=\frac{L_{0}}{L},\nonumber\\
\tilde{K_{i}}=\frac{2K_{i}}{(K_{1}+K_{3})}&;\,&\zt=\frac{z}{L}\nonumber\\
\alpha=\frac{\epsilon_{0}\Delta\epsilon(K_{1}+K_{3})}{(e_{1}-e_{3})^{2}}&;\,&{\cal E}=\frac{(e_{1}-e_{3})E}{(K_{1}+K_{3})q_{0}}.\nonumber
\end{eqnarray}
The parameter $\chi$ is a measure of the deviation of the pitch $L$ from the value implied by the intrinsic chirality $L_{0}$, while $\alpha$ is a measure of the magnitude of the reorienting effects of the dielectric anisotropy $\Delta \epsilon$.  The common liquid crystal E7 has $K_{1}=11.1$pN, $K_{2}=6.5$pN, $K_{3}=17.1$pN, $\Delta\epsilon=13.7$~\cite{Brimicombe:07} and $(e_{1}-e_{3})=12.2$pC/m~\cite{Patrick:09} giving $\Kt_{1}=0.787$, $\Kt_{2}=0.461$, $\Kt_{3}=1.213$ and $\alpha=23$.  The chiral pitch can be modified by doping, but a typical value would be $L_{0}=500$nm.

In order to respect the constraint $\n\cdot\n=1$ it is convenient to introduce an angular decomposition for the director $\n=(\ct,\st\cp,\st\sp)$ where $\ct\equiv\cos\theta$, $\sp\equiv\sin\phi$ etc.  This is the decomposition originally used by Patel and Meyer. The angle $\phi$ here represents a rotation about the $x$-axis or electric field, we will refer to this angle as the \emph{tilt} angle, while the angle $\theta$ will be referred to as the \emph{twist} angle.



\subsection{Euler-Lagrange Equations}
Using our decomposition for the director $\n=(\ct,\st\cp,\st\sp)$ the free energy in equation~\ref{eq:free} can be written:
\begin{eqnarray}
\tilde{{\cal F}}&=&\left\{\int_{\zt=0}^{\zt=1}\left[\frac{\left(g_{1}\theta_{z}^{2}+2g_{2}\theta_{z}\phi_{z}+g_{3}\phi_{z}^{2}\right)}{2}-g_{4}\theta_{z}+g_{5}\right]d\zt\right\}\nonumber\\&&+\left[g_{6}\right]_{\zt=0}^{\zt=1}\label{eq:functional},
\end{eqnarray}
where:
\begin{eqnarray}
g_{1}&=&\chi^{2}\left[(\Kt_{1}\ct^{2}+\Kt_{3}\st^{2})\sp^{2}+\Kt_{2}\cp^{2}\right]\nonumber,\\
g_{2}&=&\chi^{2}(\Kt_{1}-\Kt_{2})\sp\cp\st\ct,\nonumber\\
g_{3}&=&\chi^{2}\left[\Kt_{1}\cp^{2}\st^{2}+\sp^{2}\st^{2}(\Kt_{2}\ct^{2}+\Kt_{3}\st^{2})\right],\nonumber\\
g_{4}&=&4\pi\chi\st^{2}(\Kt_{2}\cp+{\cal E}\sp),\nonumber\\
g_{5}&=&-4\pi^{2}\alpha\ct^{2}{\cal E}^{2},\nonumber\\
g_{6}&=&2\pi\chi\st\ct(\Kt_{2}\cp+{\cal E}\sp),\nonumber
\end{eqnarray}
and $\theta_{z}\equiv d\theta/d\zt$ etc..  The equilibrium profiles for $\theta$ and $\phi$ can be deduced from the condition that the free energy in equation~\ref{eq:functional} is a minimum.  This condition gives the EL equations:

\begin{eqnarray}
g_{1}\theta_{zz}+\frac{1}{2}\frac{\partial g_{1}}{\partial \theta}\theta_{z}^{2}+\frac{\partial g_{1}}{\partial \phi}\theta_{z}\phi_{z}+\left[\frac{\partial g_{2}}{\partial \phi}-\frac{1}{2}\frac{\partial g_{3}}{\partial \theta}\right]\phi_{z}^{2}\nonumber\\-\frac{\partial g_{4}}{\partial \phi}\phi_{z}+g_{2}\phi_{zz}-\frac{\partial g_{5}}{\partial \theta}=0,\label{eq:ELtheta}
\end{eqnarray}
and
\begin{eqnarray}
g_{3}\phi_{zz}+\frac{1}{2}\frac{\partial g_{3}}{\partial \phi}\phi_{z}^{2}+\frac{\partial g_{3}}{\partial \theta}\theta_{z}\phi_{z}+\left[\frac{\partial g_{2}}{\partial \theta}-\frac{1}{2}\frac{\partial g_{1}}{\partial \phi}\right]\theta_{z}^{2}\nonumber\\
+\frac{\partial g_{4}}{\partial \phi}\theta_{z}+g_{2}\theta_{zz}=0.\label{eq:ELphi}
\end{eqnarray}
where $\theta_{zz}\equiv d^{2}\theta/d\zt^{2}$ etc.  The functions $\theta(\zt)$ and $\phi(\zt)$ which satisfy these two equations, subject to appropriate boundary conditions, are those which minimise the free energy.  We also note that since the integrand in equation~\ref{eq:functional} does not depend explicitly on the co-ordinate $\zt$ the following quantity:
\begin{equation}
g_{1}\theta_{z}^{2}+2g_{2}\theta_{z}\phi_{z}+g_{3}\phi_{z}^{2}-2g_{5}=\tau
\end{equation}
is a constant at all points $\zt$.
\subsection{Boundary Conditions}

We seek to model the situation in which the director is along $x$ for $\zt=0$ and performs one full rotation by $\zt=1$.  Thus $\n(\zt=0)=(1,0,0)$ and $\n(\zt=1)=(1,0,0)$.  In terms of the angles $\theta$ and $\phi$ at the end points these boundary conditions correspond to:  

\begin{eqnarray}
\theta(\zt=0)=0\,\, ;\,\,\theta(\zt=1)=2\pi,\\
\phi_{z}(\zt=0)=0\,\, ;\,\,\phi_{z}(\zt=1)=0.
\end{eqnarray}

The boundary conditions for $\theta(\zt)$ are trivial.  The condition for $\phi$ is somewhat more complicated.  At angles where $\sin\theta=0$ the angle $\phi$ is undefined.  A condition on the derivative of $\phi$ is therefore to be expected.  A formal proof that the correct condition has zero derivative for the angle $\phi$ is shown in appendix~\ref{sec:bcs}.  As a result of these boundary conditions, the contribution from the surface term in equation~\ref{eq:functional} is zero, and consequently we drop this term.  We note that while we do allow the angle $\phi$ to vary as a function of position in this work, our boundary conditions are consistent with a constant angle $\phi$ as imposed in previous work~\cite{Patel:87,Rudquist:94}.

\section{One Elastic Constant}\label{sec:three}

In this section we solve the EL equations for the twist and tilt angles in the special case where $K_{1}=K_{2}=K_{3}$.  For most rod-like liquid crystal molcules the elastic constants are such that $K_{3}>K_{1}>K_{2}$, however they all have similar magnitudes $K\sim 10$pN.  In this particular case the EL equations are significantly simplified and we are able to obtain analytic solutions even if we allow for pitch relaxation and dielectric behaviour.  The EL equations are now:
\begin{eqnarray}
\theta_{zz}+4\frac{\pi}{\chi}\st^{2}(\sp-{\cal E}\cp)\phi_{z}-\frac{1}{4}\st\ct\phi_{z}^{2}-8\frac{\pi^{2}}{\chi^{2}}\alpha{\cal E}^{2}\st\ct&=&0,\nonumber\\
\st\left[\st\phi_{zz}+2\ct\theta_{z}\phi_{z}-4\frac{\pi}{\chi}\st(\sp-{\cal E}\cp)\theta_{z}\right]&=&0.\nonumber\\
\end{eqnarray}
\subsection{No Dielectric Anisotropy}
If dielectric effects are unimportant ($\Delta\epsilon=0$), that is we set $\alpha=0$, the EL equations are further simplified.  The solution which satisfies these equations and the boundary conditions is then given by:
\begin{equation}
\theta(\zt)=2\pi\zt\,;\,\phi(\zt)=\text{atan}({\cal E}).\label{eq:oneconstsolnoalpha}
\end{equation}
This expression for $\phi(\zt)$ is identical to the solutions shown in equations~\ref{eq:meyer} and~\ref{eq:lee} in the one constant case.  It is interesting to note here that the expression for $\phi(\zt)$ does not depend explicitly on the pitch, i.e. we obtain the same value for $\phi(\zt)$ whether we fix the pitch to a particular value, or if we allow it to relax in order to minimise the free energy functional.  If we allow the pitch to relax, we can determine its optimal value by minimising the free energy functional with respect to (w.r.t.) $\chi$.  Substituting the solution in equation~\ref{eq:oneconstsolnoalpha} into the free energy functional in equation~\ref{eq:functional} and performing the integration over $\zt$ results in:
\begin{equation}
\tilde{{\cal F}}=2\pi^{2}\left(\chi^{2}-2\chi\sqrt{1+{\cal E}^{2}}\right).
\end{equation}
Minimising w.r.t. $\chi$ gives $\chi=\sqrt{1+{\cal E}^{2}}$. Thus if the pitch is allowed to relax we see that flexoelectric interactions lead to a reduction in the pitch (recall the pitch is proportional to $1/\chi$).

\subsection{With Dielectric Anisotropy}
If we include dielectric effects ($\Delta\epsilon\neq 0$), that is we have $\alpha\neq 0$ we still find the same solution for $\phi$, i.e. $\tan\phi={\cal E}$.  The twist angle $\theta$ then satisfies:
\begin{equation}
\chi^{2}\theta_{zz}-8\pi^{2}\alpha{\cal E}^{2}\st\ct=0,
\end{equation}
which can be integrated once to give:
\begin{equation}
\frac{(\chi\theta_{z})^{2}}{2}+4\pi^{2}\alpha{\cal E}^{2}\ct^{2}=\frac{\tau}{2},
\end{equation}
where $\tau$ is a constant.  A formal solution to this equation is given by:
\begin{equation}
\theta(\zt)=\frac{\pi}{2}-\text{Am}\left[{\cal K}_{1}(p)(1-4\zt),p\right],\label{eq:exact}
\end{equation}
where $\text{Am}[x,y]$ is the Jacobi-Amplitude function defined as the inverse of the incomplete elliptic integral of the first kind and ${\cal K}_{1}(p)$ is the complete elliptic integral of the first kind.  An identical solution was found by Davidson and Mottram~\cite{Davidson:02} (see appendix~\ref{sec:elliptic} for useful properties of the various elliptic functions).  The constant $p$ satisfies the relationship:
\begin{equation}
p{\cal K}_{1}(p)=\sqrt{\frac{\pi^{2}\alpha{\cal E}^{2}}{2\chi^{2}}}\label{eq:constraint1}
\end{equation}
and the constant $\tau=8\pi^{2}\alpha{\cal E}^{2}/p^{2}$. The constant $p$ satisfies $0\leq p\leq 1$, and we expect $p\sim\sqrt{\alpha}$ in the limit $\alpha\rightarrow 0$. The free energy per unit pitch can then be written as:
\begin{equation}
{\cal \tilde{F}}=-\frac{4\pi^{2}\alpha {\cal E}^{2}}{p^{2}}-4\chi\left(\pi^{2}\sqrt{1+{\cal E}^{2}}-\sqrt{8\pi^{2}\alpha {\cal E}^{2}}\frac{{\cal K}_{2}(p)}{p}\right),\label{eq:freeone}
\end{equation}
where ${\cal K}_{2}(p)$ is the complete elliptic integral of the second kind.  Differentiating this w.r.t. $p$ at fixed $\chi$ gives equation~\ref{eq:constraint1} as required.  We can see immediately that if we allow $\chi$ to relax rather than keeping it fixed we must have:
\begin{equation}
\pi^{2}\sqrt{1+{\cal E}^{2}}=\sqrt{8\pi^{2}\alpha {\cal E}^{2}}\left(\frac{{\cal K}_{2}(p)}{p}\right)\label{eq:constraint2}
\end{equation}
since this is the only way to have $\partial {\cal F}/\partial \chi = 0$.  The function $p{\cal K}_{1}(p)$ diverges rapidly as $p\rightarrow 1$ while ${\cal K}_{2}(p)/p$ diverges as $p\rightarrow 0$.  Taking the limit $p\rightarrow 0$, $\alpha\rightarrow 0$ in equations~\ref{eq:constraint1} and~\ref{eq:constraint2} gives:
\begin{eqnarray}
p&=&\frac{1}{2\pi\chi}\sqrt{8\pi^{2}\alpha {\cal E}^{2}}+\ldots,\label{eq:limone}\\
p&=&\frac{\sqrt{8\pi^{2}\alpha {\cal E}^{2}}}{2\pi\sqrt{1+{\cal E}^{2}}}+\ldots,
\end{eqnarray}
where we have used ${\cal K}_{1}(0)={\cal K}_{2}(0)=\pi/2$.  Comparing these two expressions for $p$ indicates that $\chi=\sqrt{1+{\cal E}^{2}}$ in the limit $\alpha=0$ as required.  For general $\alpha$ we can combine equations~\ref{eq:constraint1} and~\ref{eq:constraint2} to obtain parametric relations for ${\cal E}(\alpha,p)$ and $\chi(\alpha,p)$.
\begin{eqnarray}
{\cal E}(\alpha,p)&=&\frac{1}{\sqrt{\frac{8\alpha}{\pi^{2}}\left(\frac{{\cal K}_{2}(p)}{p}\right)^{2}-1}},\\
\chi(\alpha,p)&=&\frac{\pi}{p{\cal K}_{1}(p)}\sqrt{\frac{\alpha}{2}}\frac{1}{\sqrt{\frac{8\alpha}{\pi^{2}}\left(\frac{{\cal K}_{2}(p)}{p}\right)^{2}-1}}.
\end{eqnarray}
If $\alpha<\pi^{2}/8$ these relations produce a pitch that is a monotonically decreasing function of the applied field ${\cal E}$.  Reversing the inequality, if $\alpha>\pi^{2}/8$ the pitch initially decreases with applied field, but ultimately diverges at a critical field ${\cal E}_{c}=\pm1/\sqrt{\frac{8\alpha}{\pi^{2}}-1}$.  The divergence of the pitch corresponds to complete unwinding of the helix by the field, and in terms of dimensioned quantities occurs when:
\begin{equation}
E_{c}=\pm\left(\frac{\pi q_{0}}{2}\right)\sqrt{\frac{K}{\epsilon_{0}\Delta\epsilon-\frac{(e_{1}-e_{3})^{2}\pi^{2}}{16K}}}
\end{equation}
This expression was first obtained by Patel and Meyer~\cite{Patel:87} and is a modification to the unwinding field of a chiral nematic with no flexoelectricity and shows two important things.  Firstly the presence of flexoelectricity increases the critical unwinding field, flexoelectric coupling causes the pitch to decrease at low fields and delays the pitch divergence to higher electric fields.  Secondly for a sufficiently high ratio of the flexoelectric coefficient $(e_{1}-e_{3})$ to the dielectric anisotropy $\Delta\epsilon$ such that $(e_{1}-e_{3})^{2}/\Delta\epsilon > 16\epsilon_{0}K/\pi^{2}$ there will be no helix unwinding.
\begin{figure}
\includegraphics[width=0.4\textwidth]{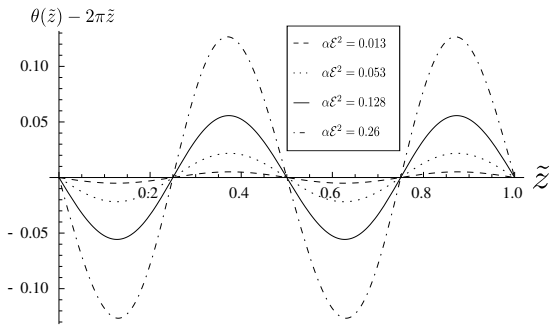}
\caption{The deviation of the twist angle from the uniform twist $\theta(\zt)-2\pi\zt$ as a function of position along the helix for several values of $\alpha{\cal E}^{2}$ for the fixed pitch case with $\chi=1$.}
\label{fig:solsoneconst}
\end{figure}
Figure~\ref{fig:solsoneconst} shows the deviation of the twist angle from the uniform state, that is $\theta(\zt)-2\pi\zt$ for several values of $\alpha {\cal E}^{2}$ with $\chi=1$.  As can be seen, the deviations grow with $\alpha {\cal E}^{2}$.  The deviations themselves appear sinusoidal, oscillating like $\sin(4\pi \zt)$.  That this should be so can be demonstrated by expanding the solution for $\theta(\zt)$ in equation~\ref{eq:exact} in terms of trigonometric functions (see appendix~\ref{sec:elliptic}), we obtain:
\begin{equation}
\theta(\zt)=2\pi \zt+\sum_{n=1}^{\infty}\left(\frac{2}{n}\right)\left(\frac{(-q)^{n}}{1+q^{2n}}\right)\sin(4n\pi \zt),
\end{equation}
from which we see that distortions from a uniform helix $\theta(\zt)=2\pi \zt$ can be represented by functions of the form $\sin(4n\pi \zt)$, i.e. a set of basis functions which satisfy the boundary conditions and are periodic on the range $\zt=0\rightarrow \zt=0.5$.

In summary, the full solution for the director structure in the one elastic constant case has $\theta(\zt)$ given by equation~\ref{eq:exact} and $\phi(\zt)=\text{atan}({\cal E})$.  If $\chi$ is fixed equation~\ref{eq:constraint1} can be used to obtain the value of $p$ needed to evaluate $\theta(\zt)$.  If $\chi$ is allowed to vary freely, equations~\ref{eq:constraint1} and~\ref{eq:constraint2} must be solved simultaneously to obtain $p$ and $\chi$ as functions of ${\cal E}$ and $\alpha$.  The value of the tilt angle $\phi$ is the same if $\chi$ is fixed or free to vary.

\section{Two Constants}\label{sec:four}
In this section we solve the EL equations for the twist and tilt angles in the special case where $K_{1}=K_{3}\neq K_{2}$ and $\Delta\epsilon=0$.  Lee~\cite{Lee:90} and Rudquist~\cite{Rudquist:94} considered this more general case.  They both assumed a constant value for $\phi$ and a uniform helix $\theta(\zt)=2\pi\zt$ and subsequently obtained the expression shown in equation~\ref{eq:lee} by minimising the resulting free energy density.  Here we demonstrate that this solution satisfies the EL equations with our more general boundary conditions for the tilt angle $\phi$.  With a constant $\phi$ the EL equations are:
\begin{eqnarray}
\chi^{2}(\sp^{2}+\Kt_{2}\cp^{2})\theta_{zz}&=&0,\nonumber\\
4\pi\chi\st^{2}\theta_{z}\left[{\cal E}\cp-\Kt_{2}\sp-\left(\frac{\theta_{z}}{2\pi}\right)\chi(1-\Kt_{2})\sp\cp\right]&=&0,\nonumber\\\label{eq:twoconstant}
\end{eqnarray}
the solutions to which are $\theta(\zt)=2\pi\zt$ and:
\begin{equation}
\tan\phi=\frac{{\cal E}}{\Kt_{2}}+\chi\frac{(\Kt_{2}-1)}{\Kt_{2}}\sin\phi.\label{eq:phitwoconstant}
\end{equation}
This equation is essentially the same as equation~\ref{eq:lee}.  However we note the presence of $\chi$ in equation~\ref{eq:phitwoconstant} which is absent in equation~\ref{eq:lee}.  The presence of this term implies that the intrinsic pitch of the cholesteric is important in determining the tilt angle even if the material is constrained to have a pitch that is different from the intrinsic value.  If we write $\delta K_{2}=1-\Kt_{2}$ we can form a series expansion for $\phi$ from equation~\ref{eq:phitwoconstant} as follows:
\begin{equation}
\phi=\text{atan}({\cal E})+\delta K_{2}\frac{{\cal E}}{(1+{\cal E}^{2})^{3/2}}(\sqrt{1+{\cal E}^{2}}-\chi)+\mathcal{O}(\delta K_{2}^{2})\label{eq:phitwoconstantexpan}
\end{equation}
Thus far we have assumed the pitch is fixed.  However we can also investigate what happens if we allow the pitch to relax.  Substituting these expressions for $\phi$ and $\theta(\zt)$ back into the free energy function in equation~\ref{eq:functional} gives:
\begin{equation}
{\cal F}=2\pi^{2}\left[(\chi\sp-{\cal E})^{2}+\Kt_{2}(\chi\cp-1)^{2}-{\cal E}^{2}-\Kt_{2}\right].
\end{equation}
Minimising we find $\chi\cp=1$ and $\chi\sp={\cal E}$ and thus:
\begin{equation}
\tan\phi={\cal E};\,\chi=\sqrt{1+{\cal E}^{2}},
\end{equation}
i.e. if we allow $\chi$ to relax we recover the one constant expression for the relaxed pitch and the tilt angle $\phi$. 

Patel and Meyer~\cite{Patel:87} also considered the case with $K_{1}=K_{3}$ and dielectric effects and with a pitch that was free to vary.  They found in this case $\tan\phi={\cal E}$, i.e. even with dielectric effects the tilt angle appeared to be spatially uniform.  We find however that if dielectric behaviour is included we can no longer find a solution of the EL equations which has $\phi$ constant in space.  The discrepancy between these two arises from the \textit{a priori} assumption of a constant $\phi$ made by Patel and Meyer.

Exact solutions to the EL equations with dielectric effects and general elastic constants are difficult to obtain.  In the next section we therefore consider a perturbation expansion about the one constant solution.

\section{Perturbation Expansions}\label{sec:foura}
In equation~\ref{eq:oneconstsolnoalpha} we obtained an exact solution for $\theta(\zt)$ and $\phi(\zt)$ in the one elastic constant case.  Solving the full EL equations with three arbitrary 
elastic constants and dielectric behaviour is difficult analytically.  An alternative we adopt here is a perturbation expansion about the solution for the one-constant case.  Thus we propose solutions
of the form:
\begin{eqnarray}
\theta(\zt)&=&2\pi \zt + \alpha \theta_{\alpha}(\zt)+\delta K_{1} \theta_{1}(\zt)+\ldots\label{eq:expanth}\\
\phi(\zt)&=&\text{atan}({\cal E})+\delta K_{1}\phi_{1}(\zt)+\delta K_{2}\phi_{2}(\zt)+\alpha\delta K_{1}\phi_{1\alpha}(\zt)+\nonumber\\
&&\alpha\delta K_{2}\phi_{2\alpha}(\zt)+\delta K_{1}^{2}\phi_{11}(\zt)+\delta K_{1}\delta K_{2}\phi_{12}(\zt)+\nonumber\\&&\delta K_{2}^{2}\phi_{22}(\zt)+\ldots\label{eq:expanphi}
\end{eqnarray}
where $\Kt_{1}=1-\delta K_{1}$\,,\,$\Kt_{3}=1+\delta K_{1}$ and $\Kt_{2}=1-\delta K_{2}$.  We note that we have not included a term  $\delta K_{2}\theta_{2}(\zt)$ in the expansion for $\theta$.  Such a term could formally be written as $\frac{\partial \theta(\zt)}{\partial \delta K_{2}}$ evaluated when $\alpha=0$, $\delta K_{1}=0$, $\delta K_{2}=0$.  However we found in section~\ref{sec:four} that $\theta(\zt)=2\pi\zt$ in the case $\alpha=0$, $\delta K_{1}=0$, i.e. $\theta_{2}(\zt)=0$.  Similarly there are no terms in the expansion for $\phi$ which depend purely on $\alpha$, since as observed in section~\ref{sec:three} the tilt angle $\phi$ does not depend on $\alpha$ if $\delta K_{1}=0$ and $\delta K_{2}=0$.  Therefore we should expect the only $\alpha$ dependent terms in $\phi$ to be mixed products of the form $\alpha\delta K_{2}$ etc.  Inserting equations~\ref{eq:expanth} and~\ref{eq:expanphi} into equation \ref{eq:ELtheta} and expanding to first order in $\delta K_{1}$ and $\alpha$:
\begin{eqnarray}
\chi^{2}\delta K_{1}\left[\theta_{1zz}+4\pi^{2}\sp^{2}\sin(4\pi \zt)\right]&&\nonumber\\
+\chi^{2}\alpha\left[\theta_{\alpha zz}-\frac{4\pi^{2}}{\chi^{2}}{\cal E}^{2}\sin(4\pi \zt)\right]&=&0.
\end{eqnarray}
Solving these equations for independent $\delta K_{1}$ and $\alpha$ subject to the boundary conditions gives:
\begin{eqnarray}
\theta_{1}(\zt)&=&\frac{{\cal E}^{2}}{4(1+{\cal E}^{2})}\sin(4\pi \zt),\nonumber\\
\theta_{\alpha}(\zt)&=&-\frac{{\cal E}^{2}}{4\chi^{2}}\sin(4\pi \zt).\label{eq:perttheta}
\end{eqnarray}
These perturbation results are accurate provided the value of the gradient of $\theta$, $\theta_{z}$, is never substantially different from $2\pi$.  This sets an upper value for the electric field for which we expect our perturbation results to be useful.  
\begin{equation}
{\cal E}^{2}<<\frac{2}{|\alpha-\delta K_{1}|}\rightarrow\left(\frac{E}{q_{0}}\right)^{2}<<\frac{2(K_{1}+K_{3})}{\left[\frac{\epsilon_{0}\Delta\epsilon}{\chi^{2}}-\frac{(K_{3}-K_{1})(e_{1}-e_{3})^{2}}{(K_{1}+K_{3})^{2}}\right]}\label{eq:pertcons}
\end{equation}
Substituting equations~\ref{eq:perttheta} and equation~\ref{eq:expanphi} into equation~\ref{eq:ELphi} and expanding to first order in $\delta K_{1}$ and $\delta K_{2}$ (not $\alpha$) we obtain:
\begin{eqnarray}
\delta K_{1} \sin(2\pi\zt)\left[\sin(2\pi \zt)(\chi \phi_{1zz}-8\pi^{2}\sqrt{1+{\cal E}^{2}}\phi_{1})\right]&&\nonumber\\
+\delta K_{1}\sin(2\pi\zt)\left[4\pi\chi\cos(2\pi \zt)\phi_{1z}\right]&&\nonumber\\
-8\pi^{2}\delta K_{2} \sin^{2}(2\pi\zt)\left[\frac{{\cal E}}{1+{\cal E}^{2}}(\chi-\sqrt{1+{\cal E}^{2}})+\sqrt{1+{\cal E}^{2}}\phi_{2}\right]&&\nonumber\\
+\delta K_{2}\sin(2\pi\zt)\left[4\pi\chi\cos(2\pi \zt)\phi_{2z}+\chi\phi_{2zz}\right]=0&&\nonumber\\
\label{eq:perturb1a}
\end{eqnarray}
the solutions to which are:
\begin{eqnarray}
\phi_{1}&=&0\\
\phi_{2}&=&\frac{{\cal E}}{(1+{\cal E}^{2})^{\frac{3}{2}}}(\sqrt{1+{\cal E}^{2}}-\chi). 
\end{eqnarray}
The term $\phi_{2}$ can be recognised as the first term of a series in powers of $\delta K_{2}$ as shown in equation~\ref{eq:phitwoconstantexpan}.  We also note that $\phi_{2}\rightarrow 0$ if $\chi=\sqrt{1+{\cal E}^{2}}$, as expected from our earlier discussions.  In fact using only the first order shifts in the angle $\theta$ we may obtain the second-order shifts in the angle $\phi$.  The general expressions for the second-order shifts are long and cumbersome, we provide them in appendix~\ref{sec:pertapp}.  For brevity we quote here the results with $\chi=1$ and expanded to cubic order in ${\cal E}$.  The expression for $\phi$ then becomes:
\begin{eqnarray}
\phi(\zt) &=& {\cal E}-\frac{{\cal E}^{3}}{3}+\frac{[3(\delta K_{1}-\alpha)\delta K_{1}+(5+\alpha-\delta K_{1})\delta K_{2}]}{10}{\cal E}^{3}\nonumber\\&&-\frac{(\delta K_{1}-2\delta K_{2})(\alpha-\delta K_{1})}{10}{\cal E}^{3}\cos(4\pi \zt).\label{eq:perturb1}
\end{eqnarray}
There are several interesting things to note about this expression.  Firstly, in general there is spatial variation in the angle $\phi$.  If however $\alpha=0$ and $\delta K_{1}=0$ the tilt angle is constant in space, but in general different from the one constant model.  This is in agreement with equation~\ref{eq:phitwoconstant}.  Furthermore we notice that at this level of perturbation, the spatial variation in $\phi$ vanishes if our liquid crystal satisfies either $\delta K_{1}=2\delta K_{2}$ or $\delta K_{1}=\alpha$.  For this later condition we also notice that the first order shifts in the angle $\theta(\zt)$ cancel (to order ${\cal E}^{3}$), i.e. a uniform helix $\theta(\zt)=2\pi\zt$ corresponds to a constant tilt angle $\phi$.  

In terms of dimensioned quantities, the condition $\delta K_{1}=2\delta K_{2}$ is equivalent to $3(K_{1}-K_{2})+(K_{3}-K_{2})=0$.  Most liquid crystals have $K_{1}>K_{2}$ and $K_{3}>K_{2}$ so this condition is not in general satisfied.  The second condition corresponds to:
\begin{equation}
(e_{1}-e_{3})^{2}=\frac{\epsilon_{0}\Delta\epsilon (K_{1}+K_{3})^{2}}{(K_{3}-K_{1})},\label{eq:delspc}
\end{equation}
for E7 this equality is not met.  For most liquid crystal materials the right hand side of this equation is somewhat larger than the left hand side, suggesting that from a material design point of view reducing the dielectric anisotropy of the materials would be useful.  This is not a new aim, several papers~\cite{Coles:01,Cole06a} have investigated the flexoelectric properties of bent-core bimesogenic compounds thought to be useful because of their low dielectric anisotropies.  What is new here is we assert that in order to ensure a uniform tilt angle $\phi$ we should want a dielectric anisotropy that is different from zero.  If we assume the E7 values for $K_{1}$, $K_{3}$ and $(e_{1}-e_{3})$ then equation~\ref{eq:delspc} rearranges to give $\Delta \epsilon=0.127$.

\section{Results}\label{sec:five}
The perturbation expansion in the previous section highlights the crucial features that occur as a result of elastic and dielectric anisotropy.  The magnitude of values for $\delta K_{1}$, $\delta K_{2}$ and $\alpha$ for common liquid crystals like E7 are such that the results of the second-order perturbation theory presented here, while qualitatively descriptive are quantitatively inaccurate.  We must thus resort to numerical solutions of the governing EL equations if we want to see what effect realistic fields will have on realistic materials.  Similar numerical modelling of the related uniformly standing helix (USH) configuration, in which the helical axis is parallel to the normal to parallel electrodes,  has been performed by Castles \textit{et. al.}~\cite{castles:09}. We first consider the situation without dielectric behaviour.  
\subsection{No Dielectric Behaviour}
\begin{figure}
\psfragscanon
\psfrag{$t1$}[Bl][Bl][0.7][0]{$\delta K_{1}=0,\delta K_{2}=0,\alpha=0.$}
\psfrag{$t2$}[Bl][Bl][0.7][0]{$\begin{array}{l}\text{x5}, \delta K_{1}=0.05,\delta K_{2}=0.1,\\\alpha=0.\end{array}$}
\psfrag{$t4$}[Bl][Bl][0.7][0]{$\delta K_{1}=0.4,\delta K_{2}=0.2,\alpha=0.$}
\psfrag{$t3$}[Bl][Bl][0.7][0]{$\delta K_{1}=0.213,\delta K_{2}=0.539,\alpha=0.$}
\psfrag{$xl$}{$\begin{array}{c}\,\\\tilde{z}\\(b)\end{array}$}
\psfrag{$yl$}{$\theta(\tilde{z})-2\pi\zt$}
\psfrag{$t1a$}[Bl][Bl][0.7][0]{$\delta K_{1}=0,\delta K_{2}=0,\alpha=0.$}
\psfrag{$t2a$}[Bl][Bl][0.7][0]{$\delta K_{1}=0.05,\delta K_{2}=0.1,\alpha=0.$}
\psfrag{$t4a$}[Bl][Bl][0.7][0]{$\delta K_{1}=0.4,\delta K_{2}=0.2,\alpha=0.$}
\psfrag{$t3a$}[Bl][Bl][0.7][0]{$\delta K_{1}=0.213,\,\delta K_{2}=0.539,\alpha=0.$}
\psfrag{$xla$}{$\begin{array}{c}\tilde{z}\\\text{(a)}\end{array}$}
\psfrag{$yla$}{$\phi(\tilde{z})$(radians)}
\includegraphics[width=0.45\textwidth]{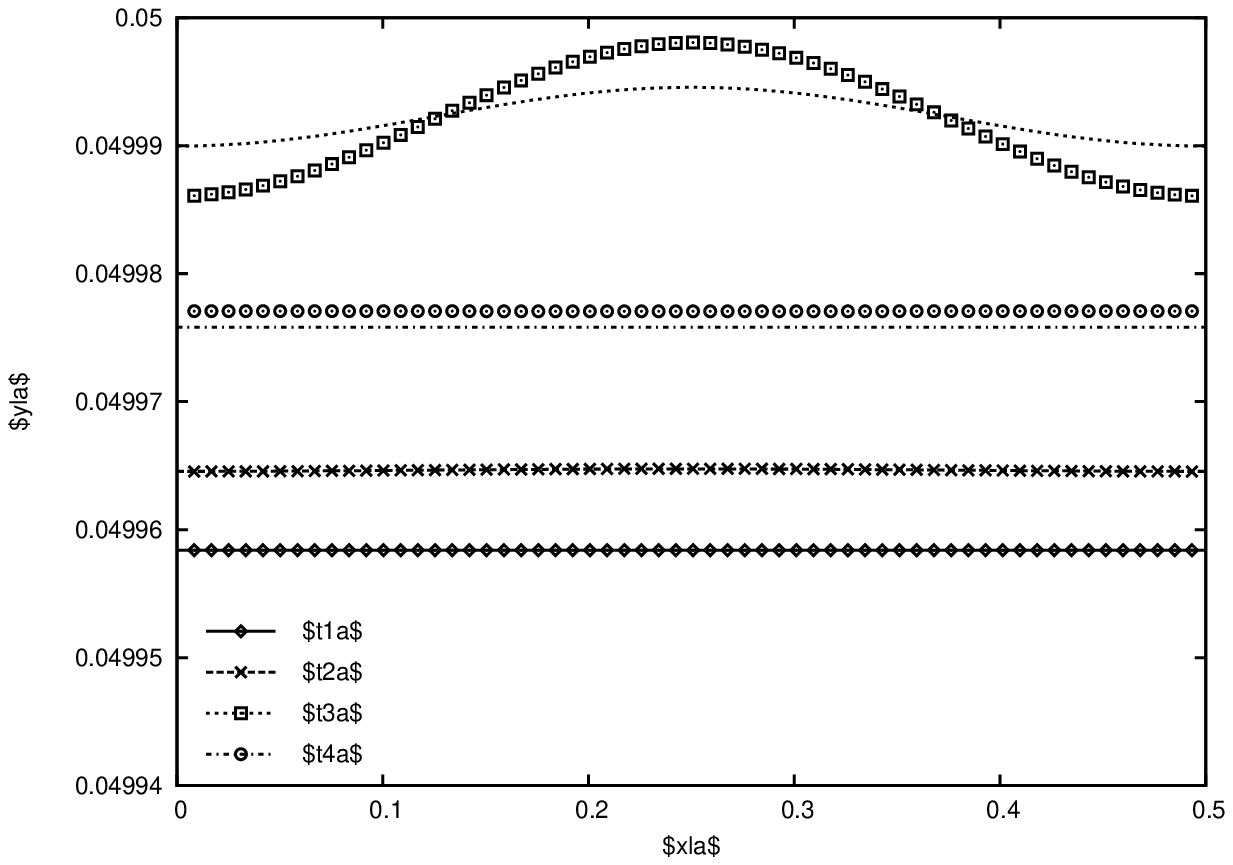}
\includegraphics[width=0.45\textwidth]{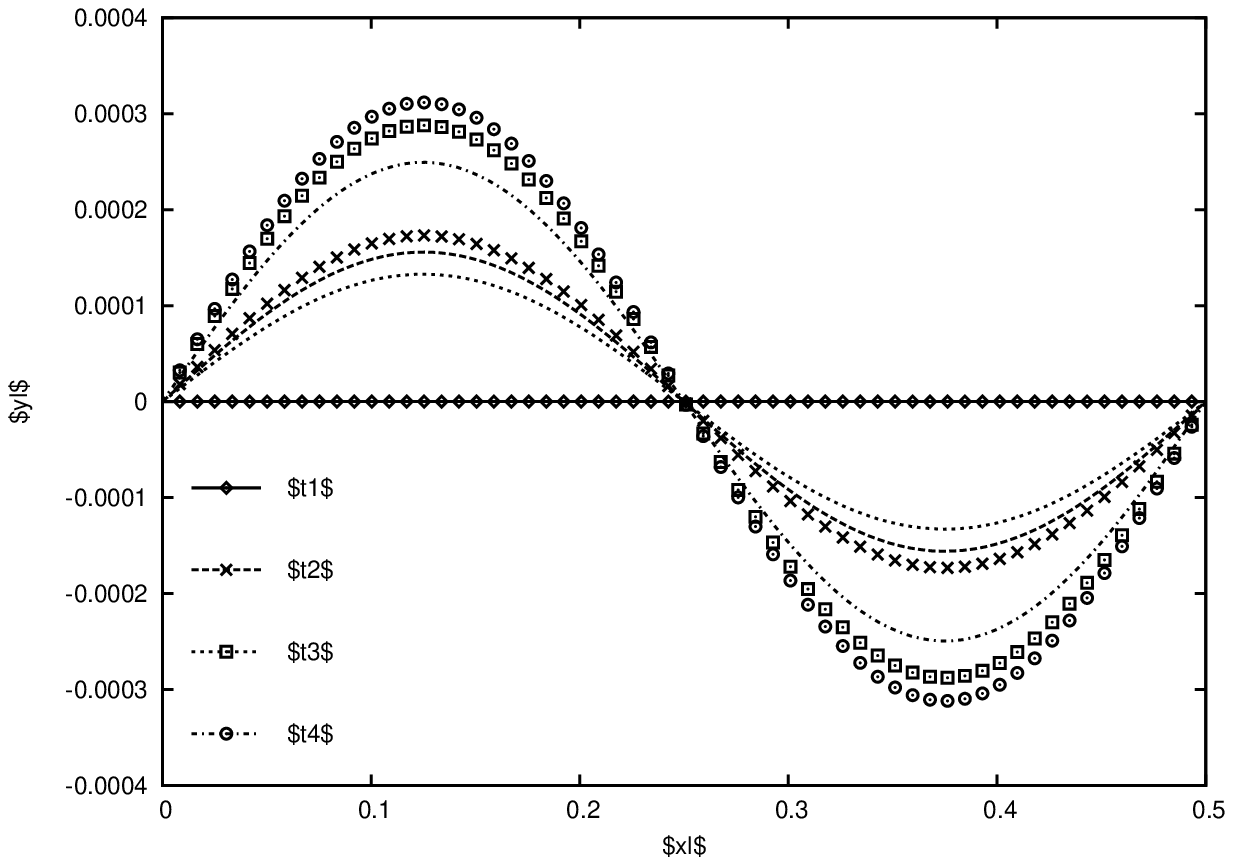}
\caption{(a) The variation of the tilt angle $\phi(\zt)$ expressed in radians as a function of $\zt$ for ${\cal E}=0.05$.  Only the first half-pitch is shown since the solutions are periodic over the range $\zt=0\rightarrow 0.5$.  Data points show numerical results for the elastic parameters indicated by the key, while lines are the corresponding perturbation approximation based on equation~\ref{eq:perturb1}. (b) The variation of the twist angle relative to a uniform helix $\theta(\zt)-2\pi\zt$ as a function of $\zt$ for ${\cal E}=0.05$.  Note in this case the data the case with $\delta K_{1}=0.05$ have been scaled by a factor of five.}
\label{fig:phinodielectric1}
\end{figure}
In figure~\ref{fig:phinodielectric1}(a) we show the variation of the tilt angle $\phi(\zt)$ as a function of position $\zt$ for ${\cal E}=0.05$.  Note we have only plotted results for the first half-pitch $\zt\leq 0.5$ since the solution is periodic over half the pitch.  The diamonds show the result for numerical minimisation in the one elastic constant case ($\delta K_{1}= \delta K_{2}=0.0$), while the line through the diamonds shows the expected analytic behaviour $\phi=\text{atan}(0.05)$.  The crosses show the result for relatively small deviations from elastic isotropy $\delta K_{1}=0.05\,,\,\delta K_{2}=0.1$.  The line passing through the crosses is based on the perturbation expansion in equation~\ref{eq:perturb1}.  At the resolution afforded by the main window in figure~\ref{fig:phinodielectric1} the agreement seems perfect.  The circles show the result for the special (and somewhat improbable) case $\delta K_{1}=0.4=2\delta K_{2}$.  As can be seen the numerical data lie slightly above the value predicted by the perturbation expansion, however the tilt angle itself appears remarkably uniform - indeed the amplitude of the variation in $\phi(\zt)$ is roughly 10$^{-5}$\% of the mean value of $\phi(\zt)$.  Finally the square data points show the numerical result for $\phi(\zt)$ for the E7 values of the elastic coefficients $\delta K_{1}=0.213\,,\,\delta K_{2}=0.539$.  As can be seen in this case there is noticeable variation in the tilt angle $\phi(\zt)$ along the helix.  The perturbation expansion captures the correct behaviour, but the amplitude of the deviations in $\phi$ is somewhat larger in the numerical data than predicted by the perturbation result.  This is to be expected since $\delta K_{2}=0.539$ is not a small deviation from elastic isotropy.  Even so, the amplitude of the distortions in $\phi(\zt)$ is only 0.024\% of the pitch-averaged value.  In figure~\ref{fig:phinodielectric1}(b) we show the twist angle relative to a unformly twisted helix, that is $\theta(\zt)-2\pi\zt$ for the same set of parameters.  Once again data points show the result of numerical minimisation, while lines show the perturbative solution based upon equation~\ref{eq:perttheta}.  Comparison of the two plots demonstrates a reciprocal relationship between $\phi(\zt)$ and the derivative of the twist angle $\theta_{z}$, i.e. $\phi(\zt)\sim 1/\theta_{z}$.  For the cases we have plotted we see that $\theta_{z}=d\theta/d\zt$ is smallest at $\zt=0.25$, that is where $\theta(\zt)=\pi/2$ and at these points the angle $\phi(\zt)$ is largest.
\begin{figure}
\psfragscanon
\psfrag{$t1$}[Bl][Bl][0.7][0]{$\delta K_{1}=0,\,\delta K_{2}=0,\,\alpha=0.$}
\psfrag{$t2$}[Bl][Bl][0.7][0]{$\begin{array}{l}\text{x5},\delta K_{1}=0.05,\delta K_{2}=0.1,\\\alpha=0.\end{array}$}
\psfrag{$t3$}[Bl][Bl][0.7][0]{$\delta K_{1}=0.4,\,\delta K_{2}=0.2,\,\alpha=0.$}
\psfrag{$t4$}[Bl][Bl][0.7][0]{$\delta K_{1}=0.213,\,\delta K_{2}=0.539,\,\alpha=0.$}
\psfrag{$xl$}{$\begin{array}{c}\,\\\tilde{z}\\(b)\end{array}$}
\psfrag{$xla$}{$\begin{array}{c}\tilde{z}\\(a)\end{array}$}
\psfrag{$yl$}{$\theta(\tilde{z})-2\pi\zt$}
\psfrag{$t1a$}[Bl][Bl][0.7][0]{$\delta K_{1}=0,\,\delta K_{2}=0,\,\alpha=0.$}
\psfrag{$t2a$}[Bl][Bl][0.7][0]{$\delta K_{1}=0.05,\,\delta K_{2}=0.1,\,\alpha=0.$}
\psfrag{$t3a$}[Bl][Bl][0.7][0]{$\delta K_{1}=0.4,\,\delta K_{2}=0.2,\,\alpha=0.$}
\psfrag{$t4a$}[Bl][Bl][0.7][0]{$\delta K_{1}=0.213,\,\delta K_{2}=0.539,\,\alpha=0.$}
\psfrag{$yla$}{$\phi(\tilde{z})$(radians)}
\includegraphics[width=0.45\textwidth]{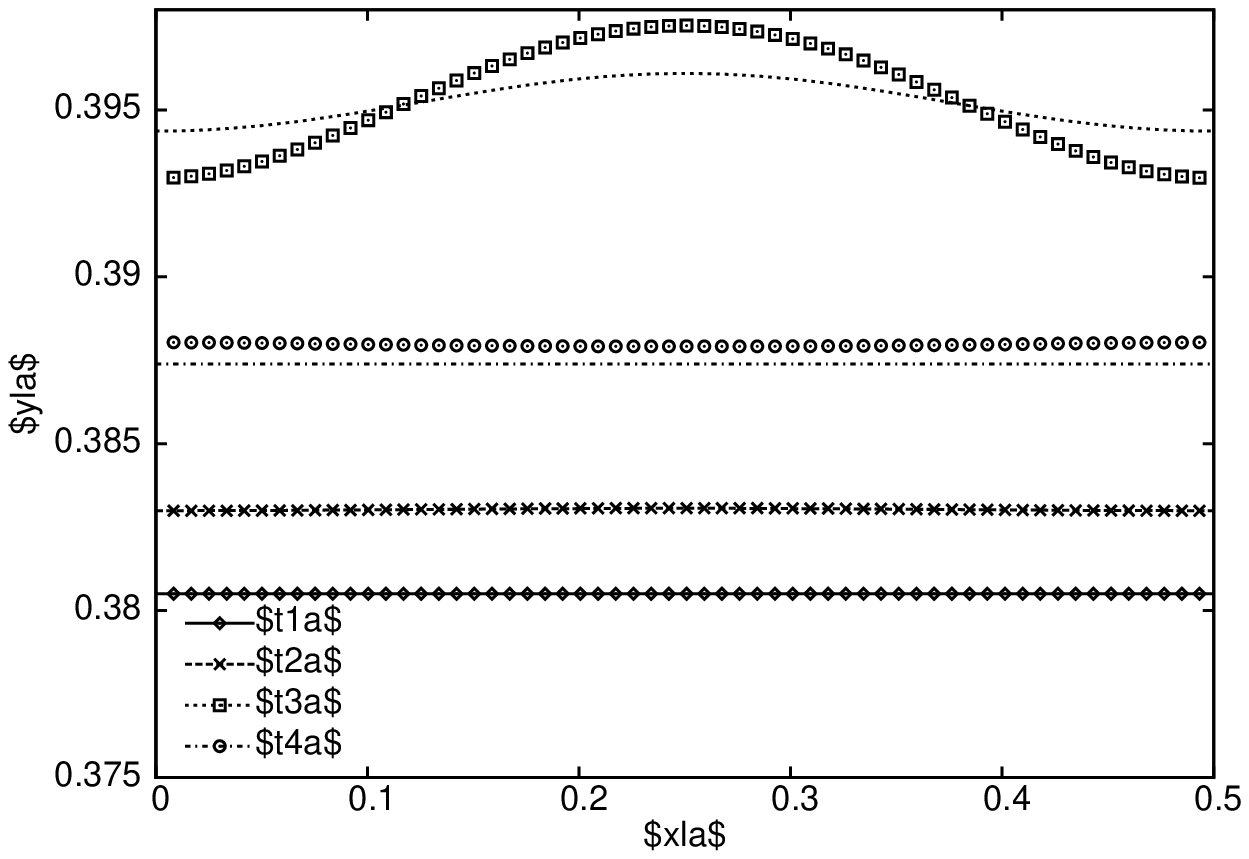}
\includegraphics[width=0.45\textwidth]{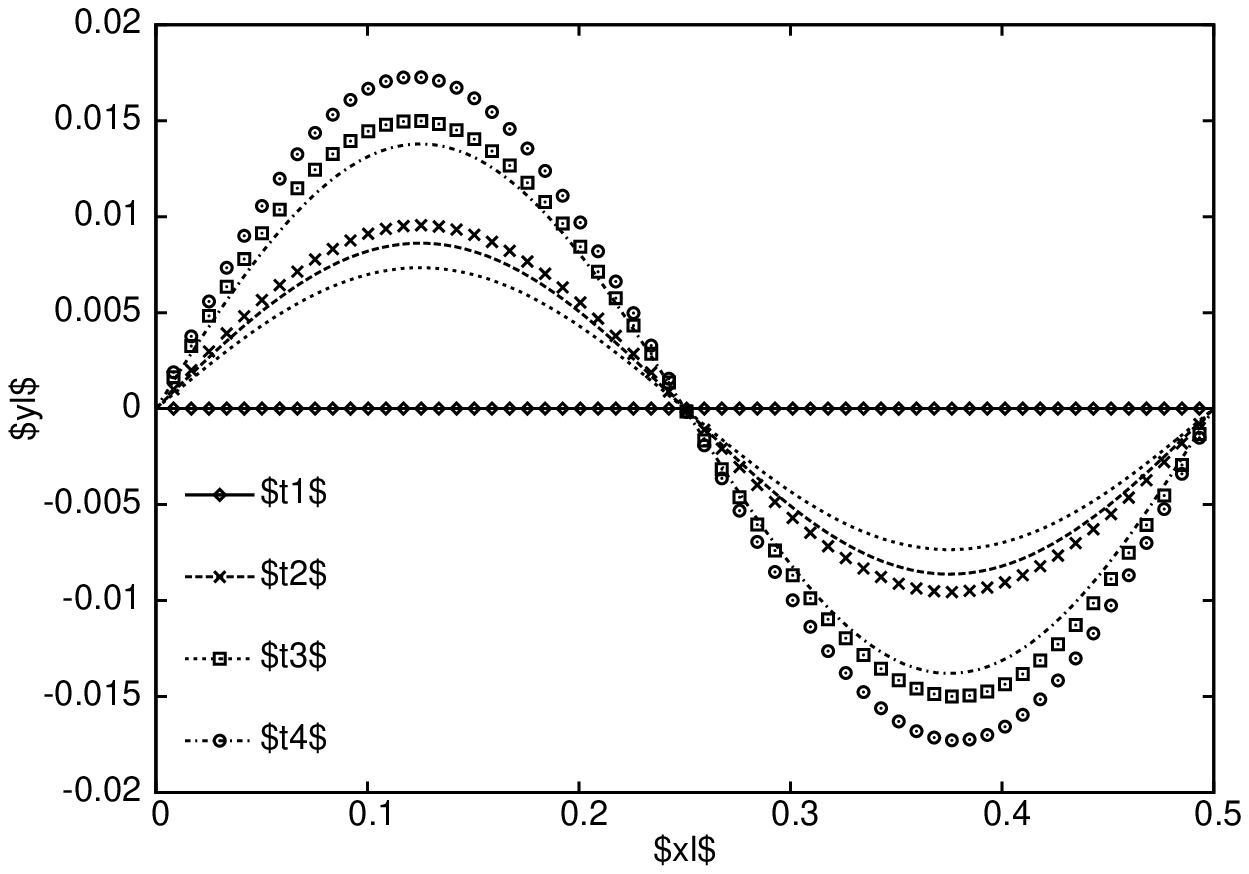}
\caption{(a) The variation of the tilt angle $\phi(\zt)$ expressed in radians as a function of $\zt$ for ${\cal E}=0.4$.  Only the first half-pitch is shown since the solutions are periodic over the range $\zt=0\rightarrow 0.5$.  Data points show numerical results for the elastic parameters indicated by the key, while lines are the corresponding perturbation approximation based on equation~\ref{eq:perturb1}.  (b) The variation of the twist angle relative to a uniform helix $\theta(\zt)-2\pi\zt$ as a function of $\zt$ for ${\cal E}=0.05$.  Note in this case the data the case with $\delta K_{1}=0.05$ have been scaled by a factor of five.}
\label{fig:phinodielectric2}
\end{figure}

In figure~\ref{fig:phinodielectric2} we show similar plots for the angle $\phi(\zt)$ and $\theta(\zt)$, however this time for ${\cal E}=0.4$.  This choice of ${\cal E}$ should give a value of $\phi\sim 22.5^{\circ}~\sim 0.4$ radians.  This choice is motivated by device applications.  A ULH is usually modelled as a uniaxial slab placed between crossed polarisers.  If $\varphi$ is the angle between one of the polarisers and the unique axis of the slab the transmission of light intensity should vary as $\sin^{2}(2\varphi)$.  Thus in order to get full intensity modulation we need to be able to change $\varphi$ by $\pi/4$.  This is satisfied if we set $\varphi=\pi/8+\phi$, provided that we are able to produce flexoelectric tilts $\phi$ of $\pm \pi/8\equiv\pm 22.5^{\circ}$.  The diamonds show the numerical data in the one constant case, this agrees with the expected analytic value  $\phi=\text{atan}(0.4)=0.381$ radians.  The crosses show the numerical result for $\phi(\zt)$ when $\delta K_{1}=0.05,\,\delta K_{2}=0.1$ while the line shows the perturbative result.  Once again the circles show the result for the case $\delta K_{1}=0.4=2\delta K_{2}$.  The numerical data lie slightly above the value predicted by the perturbative result, but the tilt angle itself is highly uniform.  The peak-to-peak amplitude of the variation in $\phi(\zt)$ is 0.034\% relative to the mean value of $\phi$.  The square data points show the expected behaviour when the values for the elastic constants for E7 are used.  There is observable variation in the tilt angle $\phi(\zt)$, however the amplitude of the variation in $\phi(\zt)$ is still only 0.58\% of the mean value. Once again we notice a reciprocal relationship between $\phi$ and $\theta_{z}$.

In summary the plots for the E7 elastic parameters are perhaps the most useful here.  We have demonstrated that the tilt angle $\phi(\zt)$ does in general vary with position.  The exceptions to this being when $\delta K_{1}=0$ or $\delta K_{1}=2\delta K_{2}$.  However we have also demonstrated that for a material with the E7 elastic constants, in the absence of dielectric behaviour the magnitude of the distortion in the angle $\phi(\zt)$ is roughly 0.6\% of the mean value for the physically motivated value of $\phi=22.5^{\circ}$.  We can thus conclude that current understanding of the helical-chiral-flexo-electro optic effect based on equations~\ref{eq:meyer} and~\ref{eq:lee}, while not founded on wholly accurate assumptions, does provide an accurate description of the switching behaviour.

\subsection{Dielectric Behaviour}
\begin{figure}[!h]
\psfragscanon
\psfrag{$t1$}[Bl][Bl][0.7][0]{x20 $\delta K_{1}=0.213,\,\delta K_{2}=0.539,\,\alpha=0.$}
\psfrag{$t2$}[Bl][Bl][0.7][0]{$\begin{array}{l}\text{x500}, \delta K_{1}=0.213,\delta K_{2}=0.539,\\\alpha=0.213\end{array}$}
\psfrag{$t3$}[Bl][Bl][0.7][0]{$\delta K_{1}=0.213,\,\delta K_{2}=0.539,\,\alpha=23$}
\psfrag{$t4$}[Bl][Bl][0.7][0]{$\delta K_{1}=0.213,\,\delta K_{2}=0.539,\,\alpha=0.$}
\psfrag{$xl$}{$\begin{array}{c}\,\\\tilde{z}\\(b)\end{array}$}
\psfrag{$xla$}{$\begin{array}{c}\tilde{z}\\(a)\end{array}$}
\psfrag{$yl$}{$\theta(\tilde{z})-2\pi\zt$}
\psfrag{$t1a$}[Bl][Bl][0.7][0]{$\delta K_{1}=0.213,\,\delta K_{2}=0.539,\,\alpha=0.$}
\psfrag{$t2a$}[Bl][Bl][0.7][0]{$\delta K_{1}=0.213,\,\delta K_{2}=0.539,\,\alpha=0.213$}
\psfrag{$t3a$}[Bl][Bl][0.7][0]{$\delta K_{1}=0.213,\,\delta K_{2}=0.539,\,\alpha=23$}
\psfrag{$t4a$}[Bl][Bl][0.7][0]{$\delta K_{1}=0.213,\,\delta K_{2}=0.539,\,\alpha=0.$}
\psfrag{$yla$}{$\phi(\tilde{z})$(radians)}
\includegraphics[width=0.5\textwidth]{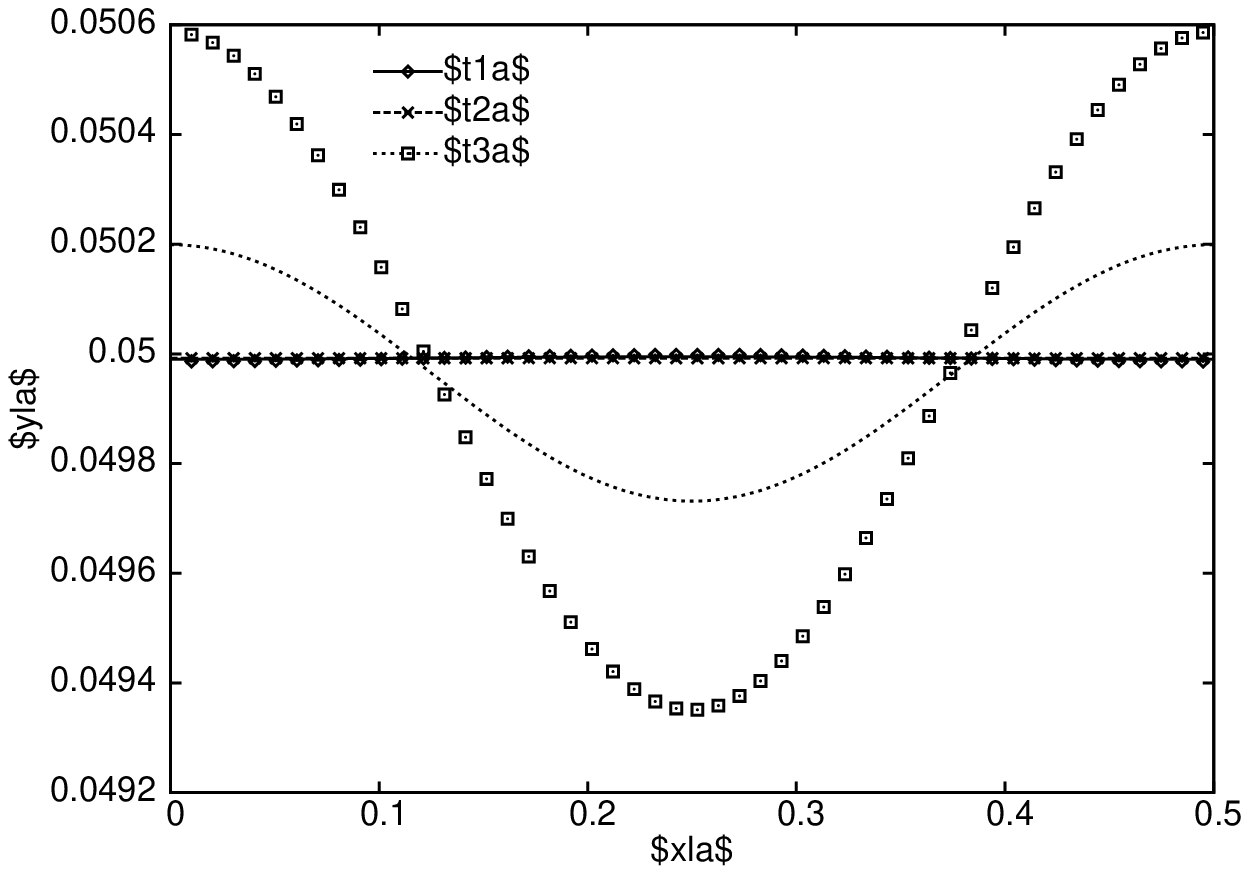}
\includegraphics[width=0.5\textwidth]{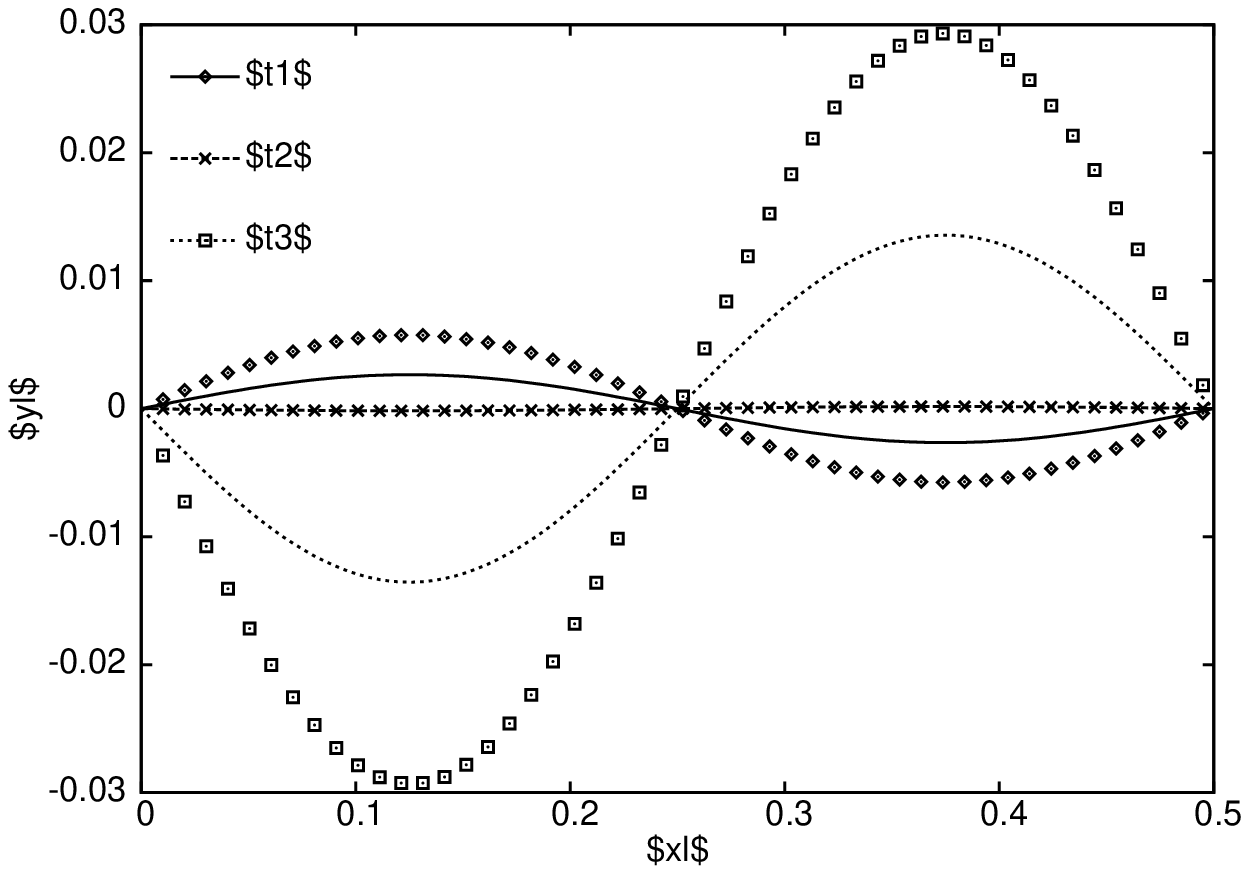}
\caption{(a) The variation of the tilt angle $\phi(\zt)$ expressed in radians as a function of $\zt$ for ${\cal E}=0.05$.  Only the first half-pitch is shown since the solutions are periodic over the range $\zt=0\rightarrow 0.5$.  Data points show numerical results for the elastic parameters indicated by the key, while lines are the corresponding perturbation approximation based on equation~\ref{eq:perturb1}.  (b) The variation of the twist angle relative to a uniform helix $\theta(\zt)-2\pi\zt$ as a function of $\zt$ for ${\cal E}=0.05$.  Note in this case the data the case with $\alpha=0$ have been scaled by a factor of 20 and those with $\alpha=0.213$ by a factor of 500.}
\label{fig:phidielectric1}
\end{figure}
We now include dielectric effects in order to determine how they alter the behaviour of $\phi(\zt)$.  Figure~\ref{fig:phidielectric1}(a) shows the angle $\phi(\zt)$ for a material with the E7 elastic constants and several different values of $\alpha$ all with ${\cal E}=0.05$.  The diamond ticks show numerical data for $\alpha=0$.  These agree with the square data points shown in figure~\ref{fig:phinodielectric1}.  The crosses correspond to the special case $\alpha=\delta K_{1}$.  We saw from the perturbation expansion in equation~\ref{eq:perturb1} that for this case we expect essentially no spatial variation for the tilt angle $\phi(\zt)$.  The square data points show the result for $\alpha=23$, the true value for E7.  In this case we notice there is observable variation in the tilt angle $\phi$, nevertheless it is still relatively small, the amplitude of the variation represents 1.2\% of the mean value of $\phi$.  In figure~\ref{fig:phidielectric1}(b) we show similar plots of $\theta(\zt)-2\pi\zt$.  In this case we have exaggerated the magnitude of the distortions for the $\alpha=0$ case by a factor of 20 and for the $\alpha=0.213$ case by a factor of 500.  It is interesting to note that even with a magnification of 500 the twist angle appears to describe a perfect helix for the $\alpha=0.213$ case.  It is also interesting to note how the form of $\theta(\zt)-2\pi\zt$ changes as we increase $\alpha$.  For $\alpha=0$ (diamond data-points) we see that $\theta_{z}$ is large near $\zt=0$ and $\zt=0.5$ where the director is parallel to the field. While $\theta_{z}$ is small for $\zt=0.25$ when the director is perpendicular to the field.  On the other hand for large $\alpha$ (square data-points) we see the opposite behaviour, $\theta_{z}$ is large around $\zt=0.25$ and small around $\zt=0$ and $\zt=0.5$ - the director tends to align with the field in this case.  For the special case $\alpha=\delta K_{1}$ these two effects largely cancel and we are left with a uniform helix.
\begin{figure}[!h]
\psfragscanon
\psfrag{$t1$}[Bl][Bl][0.7][0]{x20 $\delta K_{1}=0.213,\delta K_{2}=0.539,\alpha=0.$}
\psfrag{$t2$}[Bl][Bl][0.7][0]{$\begin{array}{l}{\text x500},\delta K_{1}=0.213,\delta K_{2}=0.539,\\\alpha=0.213\end{array}$}
\psfrag{$t3$}[Bl][Bl][0.7][0]{$\begin{array}{l}\delta K_{1}=0.213,\,\delta K_{2}=0.539,\\\alpha=0.539\end{array}$}
\psfrag{$t4$}[Bl][Bl][0.7][0]{$\delta K_{1}=0.213,\,\delta K_{2}=0.539,\,\alpha=0.$}
\psfrag{$yl$}{$\theta(\tilde{z})-2\pi\zt$}
\psfrag{$t1a$}[Bl][Bl][0.7][0]{$\delta K_{1}=0.213,\delta K_{2}=0.539,\alpha=0.$}
\psfrag{$t2a$}[Bl][Bl][0.7][0]{$\delta K_{1}=0.213,\delta K_{2}=0.539,\alpha=0.213$}
\psfrag{$t3a$}[Bl][Bl][0.7][0]{$\delta K_{1}=0.213,\delta K_{2}=0.539,\alpha=23$}
\psfrag{$t4a$}[Bl][Bl][0.7][0]{$\delta K_{1}=0.213,\delta K_{2}=0.539,\alpha=0.$}
\psfrag{$xla$}{$\begin{array}{c}\tilde{z}\\(a)\end{array}$}
\psfrag{$yla$}{$\phi(\tilde{z})$(radians)}
\psfrag{$xl$}{$\begin{array}{c}\,\\\tilde{z}\\(b)\end{array}$}
\includegraphics[width=0.5\textwidth]{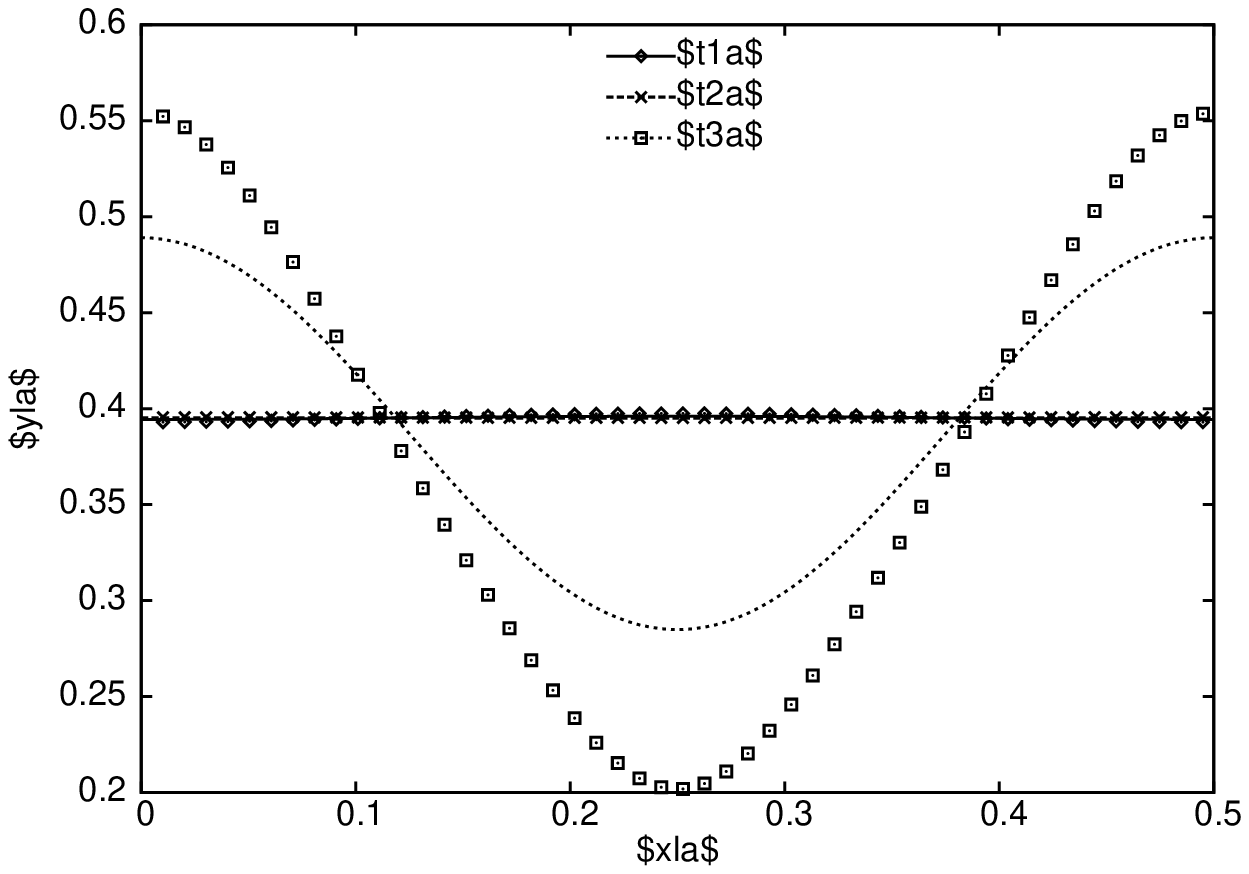}
\includegraphics[width=0.5\textwidth]{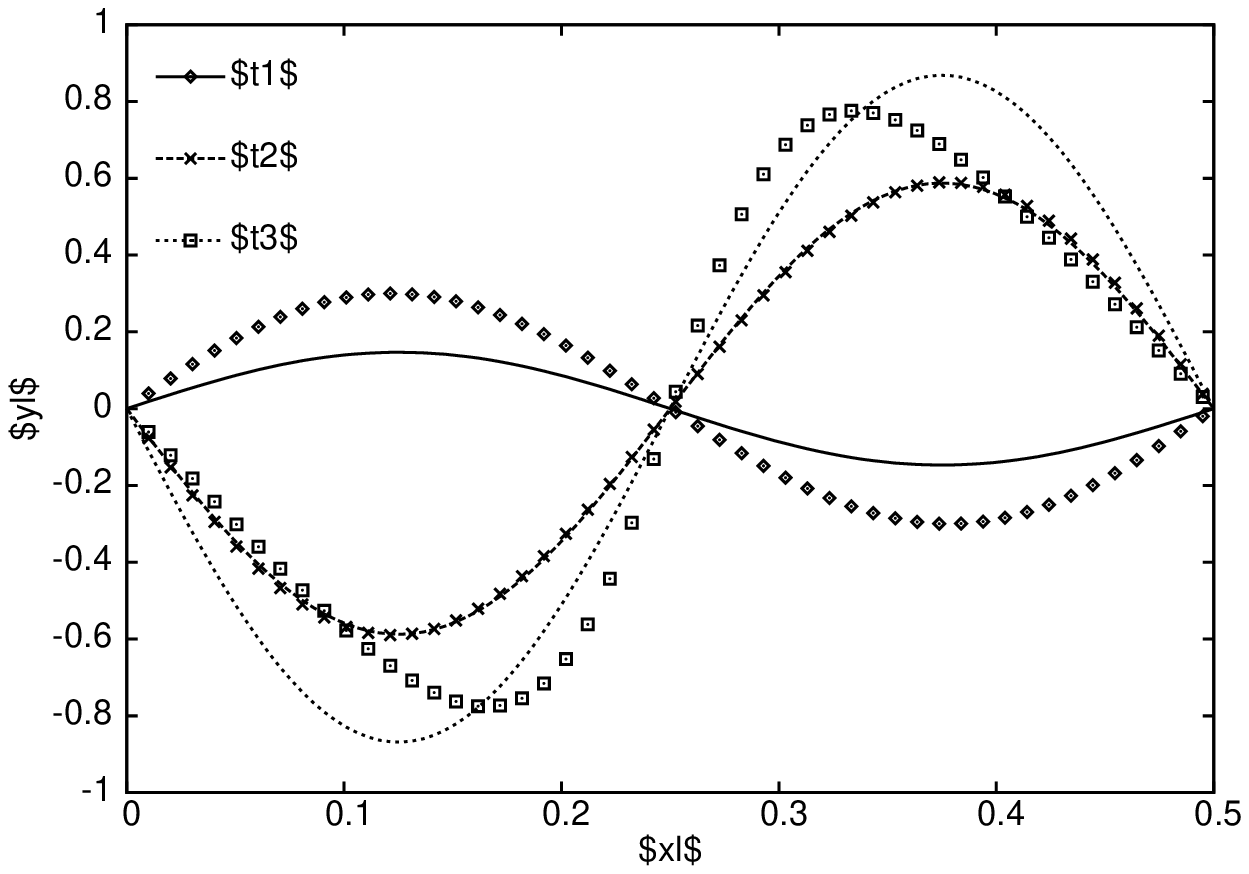}
\caption{(a) The variation of the tilt angle $\phi(\zt)$ expressed in radians as a function of $\zt$ for ${\cal E}=0.4$.  Only the first half-pitch is shown since the solutions are periodic over the range $\zt=0\rightarrow 0.5$.  Data points show numerical results for the elastic parameters indicated by the key, while lines are the corresponding perturbation approximation based on equation~\ref{eq:perturb1}.  (b) The variation of the twist angle relative to a uniform helix $\theta(\zt)-2\pi\zt$ as a function of $\zt$ for ${\cal E}=0.4$.  Note in this case the data the case with $\alpha=0$ have been scaled by a factor of 20 and those with $\alpha=0.213$ by a factor of 500.}
\label{fig:phidielectric2}
\end{figure}
In figure~\ref{fig:phidielectric2} we show similar plots but this time for ${\cal E}=0.4$.  For the $\alpha=0$ and $\alpha=\delta K_{1}=0.213$ data we again see a largely uniform $\phi(\zt)$.  The data for $\alpha=23$ however show a highly non-uniform $\phi(\zt)$ - the amplitude of the variation in $\phi$ is $\sim$40\% of the mean value.  Such a large variation is clearly at odds with the usual analytic models which predict a uniform tilt.  It is worth considering the implications this plot has for the optics of a ULH device.  

\begin{figure}
\psfrag{$t1$}[Bl][Bl][0.55][0]{$\phi(\theta=0)$}
\psfrag{$t2$}[Bl][Bl][0.55][0]{$\phi(\theta=\pi/2)$}
\psfrag{$t3$}[Bl][Bl][0.55][0]{$(\phi(\theta=0)+\phi(\theta=\pi/2))/2$}
\psfrag{$t4$}[Bl][Bl][0.55][0]{$\tan\phi=\frac{{\cal E}}{\Kt_{2}}+\frac{(\Kt_{2}-1)}{\Kt_{2}}\sin\phi$}
\psfrag{$xl$}{${\cal E}$}
\psfrag{$xl$}{$\begin{array}{r}\,\\{\cal E}\\(b)\end{array}$}
\psfrag{$xla$}{$\begin{array}{r}{\cal E}\\(a)\end{array}$}
\psfrag{$yl$}{$\phi$}
\includegraphics[width=0.45\textwidth]{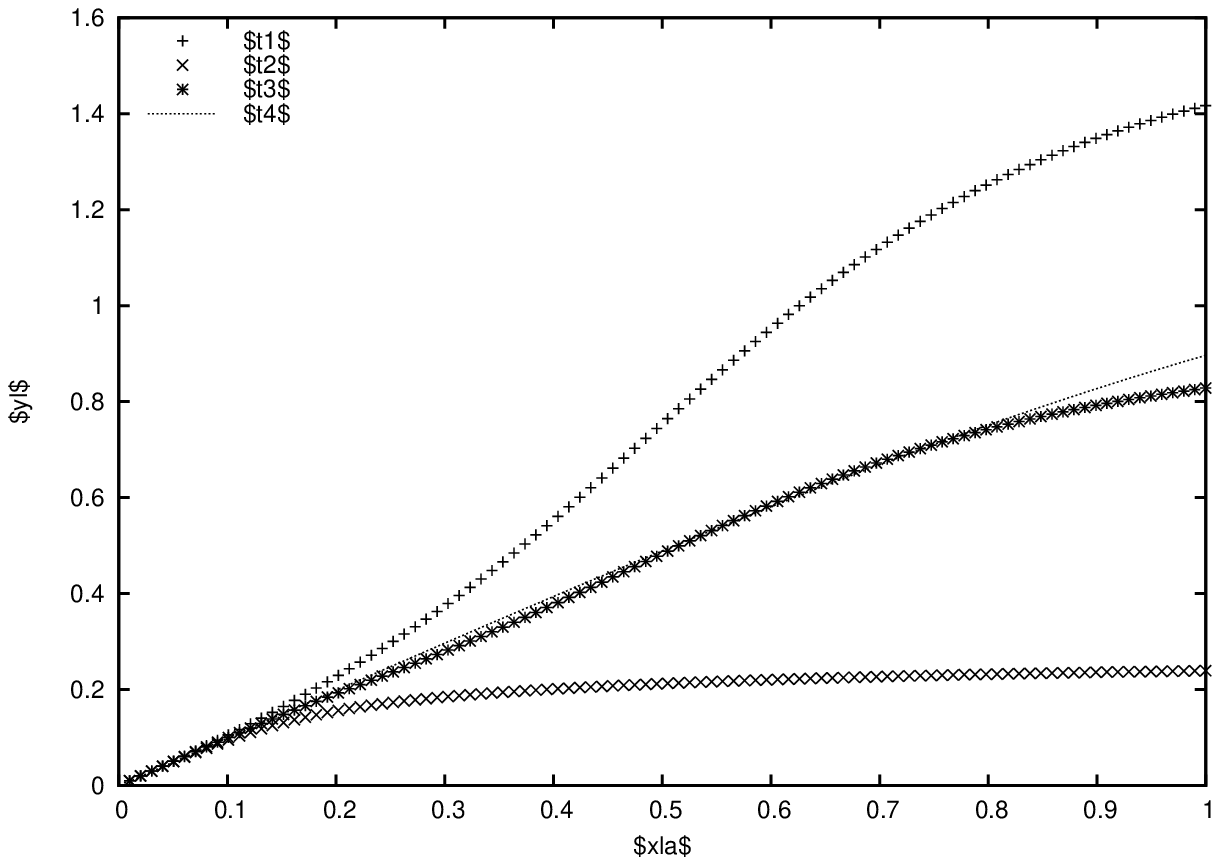}
\includegraphics[width=0.45\textwidth]{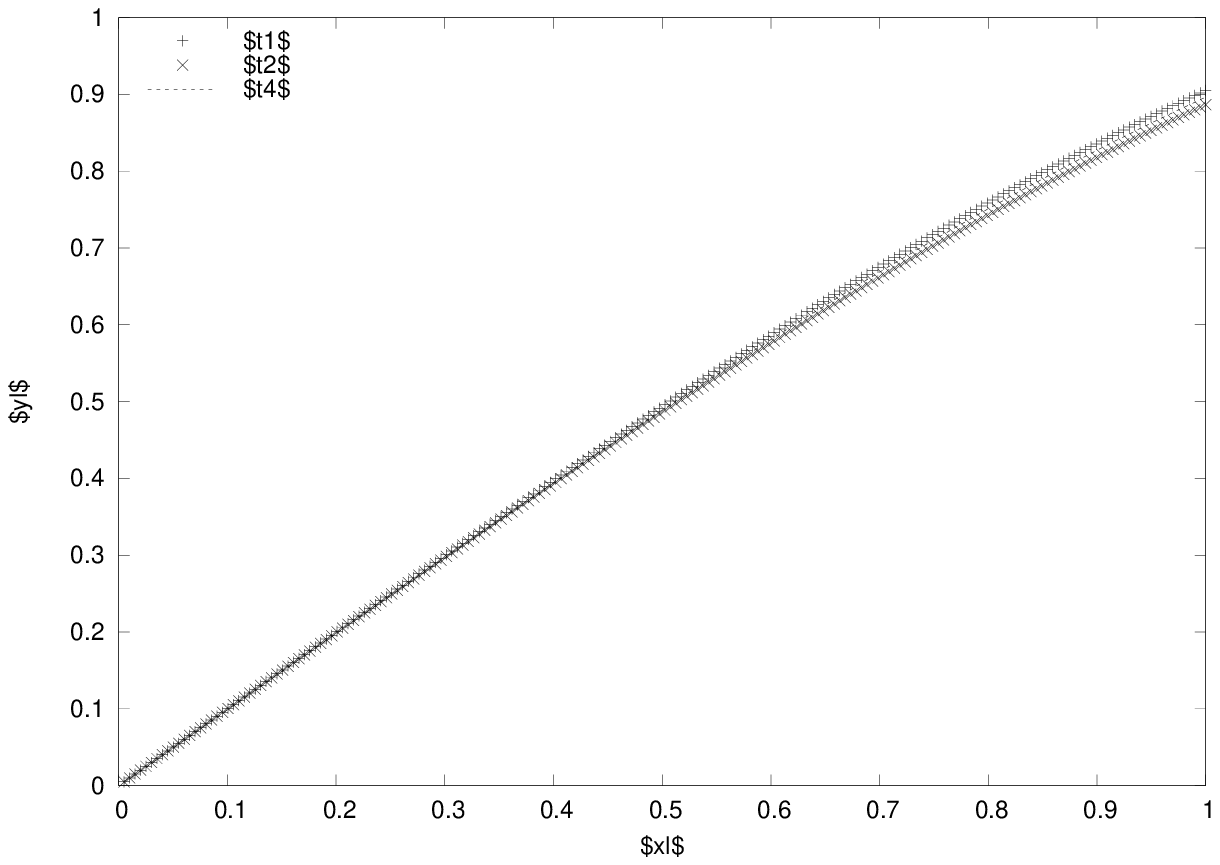}
\caption{(a) A plot showing the value of $\phi$ when $\theta=0$ (maximum) and $\theta=\pi/2$ (minimum).  Also shown is the arithmetic mean of these two and the solution for $\phi$ based upon equation~\ref{eq:phitwoconstant}.  The plot is for a material with the E7 values for the elastic and dielectric constants. (b) A similar plot however this time the material has the E7 elastic constant, but no dielectric anisotropy.}
\label{fig:elston}
\end{figure}

At $\zt=0$ the director is aligned along the electric field, for a ULH device this corresponds to the director being parallel to the normal to the electrodes, while at $\zt=1/4$ we have $\theta=\pi/2$ and the director is in the plane of the device.  A ULH is usually modelled optically as a uniaxial slab with unique axis in the plane of the electrodes and thus perpendicular to the field. The helical-flexo-electro-optic effect is then thought to rotate the unique axis around the field direction.  For a constant angle $\phi$ it is clear how much the uniaxial direction rotates by.  However, we have demonstrated that there are quite large differences between $\phi(\theta=0)$ and $\phi(\theta=\pi/2)$ if dielectric effects are taken into account.  The birefringence properties of the ULH are principally determined by the points where $\theta=\pi/2$, i.e. when the director is in the plane of the ULH device.  We have shown that at larger values of ${\cal E}$ these points have an angle $\phi$ associated with them which is substantially smaller than would be predicted by previous analytical models based on equations~\ref{eq:meyer} and~\ref{eq:phitwoconstant}, i.e. for the data for E7 shown in figure~\ref{fig:phidielectric2}(a) $\phi(\theta=\pi/2)\approx 0.2$ when we would predict $\phi(\theta(\pi/2)\approx 0.4$.  To investigate this effect further we show in figure~\ref{fig:elston}(a) a plot of $\phi(\theta=0)$ and $\phi(\theta=\pi/2)$ as well as their average and a plot of $\phi$ based upon equation~\ref{eq:phitwoconstant} against ${\cal E}$ for the E7 material parameters.  We can see that the average value of $\phi$ is modelled quite well by equation~\ref{eq:phitwoconstant} but the difference between $\phi(\theta=0)$ and $\phi(\theta=\pi/2)$ becomes large with increasing ${\cal E}$, and in particular $\phi(\theta=\pi/2)$ becomes a slowly increasing function of ${\cal E}$.  In figure~\ref{fig:elston}(b) we show a similar plot with the dielectric anisotropy set to zero, but maintaining the E7 elastic parameters.  We now see that equation~\ref{eq:phitwoconstant} models the data well over a large range of ${\cal E}$.

In summary we have shown that dielectric effects can lead to substantial variation in the tilt angle $\phi(\zt)$ as one progresses along the helix.  As the electric field is increased the tilt angle obtained in the plane of a ULH device becomes substantially smaller than one would expect based on previous modelling.  This again suggests the need for materials with low dielectric anisotropy.  In fact we might also consider negative dielectric anisotropy materials, we noted that for positive dielectric anisotropy materials the tilt angle $\phi$ is largest when the director is aligned with the field and smallest when the director is perpendicular to the field.  Changing to a negative dielectric anisotropy material should result in the a larger tilt angle perpendicular to the field.
\section{Conclusion}\label{sec:six}
In this work we have relaxed several of the assumptions previously applied in analytical models of the helical-flexo-electro-optic effect.  We have found that allowing for general elastic constants rather than restricting to special cases such as the one elastic constant model results in some spatial variation of the tilt angle $\phi$.  Nevertheless this spatial variation seems small for realistic elastic parameters provided there is no dielectric anisotropy.  In these cases while previous analytical modelling is not rigoursly correct, it is still highly accurate, at least up to the fields required to produce $\pi/8$ switching angles.

If we include dielectric anisotropy we have shown that there can be substantial spatial variation in the tilt angle $\phi$ and that deviations of the extremal values of the tilt angle can be large when compared with the value predicted by previous analytical models.  The optical modelling of the ULH usually assumes a constant tilt angle $\phi$, the large spatial distortions we report for $\phi(\zt)$ suggest more careful modelling of the transmission properties of the ULH is required.   Furthermore the large distortions in $\phi$ we report for E7 suggest the use of low dielectric anisotropy materials in order to achieve a relatively uniform tilt angle.  It is possible that using negative dielectric anisotropy materials might be a useful way to enhance the tilt angles achieved in the plane of a ULH device, and thus allow larger switching angles for lower fields.

Our perturbation analysis has highlighted several special relations between the elastic constants, flexo-electric constants and dielectric anisotropy which tend to lead to a spatially uniform tilt angle.  

In future it is hoped to extend this work by including dynamical processes (allowing for flow) and extending the structure modelling to two dimenesions (allowing for surface interactions).  Additionally more detailed modelling of the optical properties of the helical-flexo-electro-optic effect will be undertaken.
\section{Acknowledgements}
The authors would like to thank Dr Philip Benzie, Dr Giovanni Carbonne, Dr Flynn Castles and Dr Patrick Salter for useful discussions.  Funding was provided by the EPSRC under Grant No. EP/F013787/1. 
\appendix
\section{Boundary Conditions}\label{sec:bcs}

In this article we have decomposed the director $\n=(n_{x}(\zt),n_{y}(\zt),n_{z}(\zt))$ in terms of two angles $\n=(\cos\theta,\sin\theta\cos\phi,\sin\theta\sin\phi)$.  We have applied the boundary conditions $\n(\zt=0)=\n(\zt=1)=(1,0,0)$, which gives $\theta(\zt=0)=0$ and $\theta(\zt=1)=2\pi$.  The angle $\phi$ however is undefined at these points.  The tangent of the angle $\phi$ is given by:
\begin{equation}
\tan\phi(\zt)=\frac{n_{z}(\zt)}{n_{y}(\zt)}.
\end{equation}
Denoting differentiation w.r.t. $\zt$ by a prime the derivative of $\phi(\zt)$ is given by:
\begin{equation}
\phi'(\zt)=\frac{n_{y} n_{z}'-n_{z} n_{y}'}{n_{y}^{2}+n_{z}^{2}}.
\end{equation}
At the points $\zt=0$ and $\zt=1$ both the numerator and the denominator in this expression are zero since $n_y$ and $n_z$ are zero at these points.  Applying L'H\^{o}pital's rule to both numerator and denominator we arrive at:
\begin{equation}
\phi'(\zt=0/1)=\frac{(n_{y}'n_{z}''-n_{z}'n_{y}'')}{2(n_{y}^{'2}+n_{z}^{'2})}.\label{eq:phicondition}
\end{equation}
In terms of the director components the equilibrium equations are $\uline{h}=\lambda \n$ where $\uline{h}$ is the molecular field and is related to the free energy density $f$ of equation~\ref{eq:freenonscaled} by:
\begin{equation}
h_{p}=\frac{\partial f}{\partial n_{p}}-\frac{d}{d\zt}\left(\frac{\partial f}{\partial \left[\frac{dn_{p}}{d\zt}\right]}\right),
\end{equation}
and $\lambda$ is a Lagrange multiplier that constrains the director to have unit magnitude.  Evaluating the molecular field at points where the director is given by $\n=(1,0,0)$, and recalling that since the director has unit magnitude $n_{x}'=0$ at these points, we obtain the following for $h_{y}$ and $h_{z}$:
\begin{eqnarray}
h_{y}&=&K_{2}n_{y}''=0,\\
h_{z}&=&K_{1}n_{z}''=0.
\end{eqnarray}
Inserting these two relations into equation~\ref{eq:phicondition} we see that the condition on the angle $\phi$ corresponding to demanding the director satisfies $\n=(1,0,0)$ is thus $\phi_{z}=0$.
\section{Elliptic Functions}
\label{sec:elliptic}
There are several conventions used to define the elliptic functions, the definitions used in this paper are presented here.  The complete elliptic function of the first kind ${\cal K}_{1}(p)$ is defined by the integral:
\begin{equation}
{\cal K}_{1}(p)=\int_{0}^{\pi/2}\frac{1}{\sqrt{1-p^{2}\st^{2}}}d\theta,
\end{equation}
where $p^{2}<1$.  For $p\rightarrow 0$ the complete elliptic of the first kind is approximately given by:
\begin{equation}
{\cal K}_{1}(p)\approx\frac{\pi}{2}\left[1+\frac{p^{2}}{4}+\ldots\right],
\end{equation}y
while for $p\rightarrow 1$ a useful asymptotic form for the ${\cal K}_{1}(p)$ is:
\begin{equation}
{\cal K}_{1}(p)\approx\log\left[\frac{4}{\sqrt{1-p^{2}}}\right].
\end{equation}
The complete elliptic function of the second kind is defined by:
\begin{equation}
{\cal K}_{2}(p)=\int_{0}^{\pi/2}\sqrt{1-p^{2}\st^{2}}d\theta.
\end{equation}
The derivatives of the functions satisfy:
\begin{eqnarray}
\frac{d{\cal K}_{2}}{dp}&=&\frac{{\cal K}_{2}(p^{2})-{\cal K}_{1}(p^{2})}{p},\\
\frac{d{\cal K}_{1}}{dp}&=&\frac{{\cal K}_{2}(p^{2})}{p(1-p^{2})}-\frac{{\cal K}_{1}(p^{2})}{p}.
\end{eqnarray}
The incomplete elliptic integral of the first kind is given by:
\begin{equation}
{\cal F}(\phi,p)=\int_{0}^{\phi}\frac{1}{\sqrt{1-p^{2}\st^{2}}}d\theta,
\end{equation}
The Jacobi-Amplitude function $\text{Am}({\cal F}(\phi,p),p)$ is defined as the inverse of the incomplete elliptic integral.  I.e given a value for ${\cal F}(\phi,p)$ and $p$ the Jacobi-Amplitude function gives the corresponding value of $\phi$.  A useful expansion for the Jacobi-Amplitude function is:
\begin{equation}
\text{Am}(u,p)=\frac{\pi u}{2{\cal K}_{1}(p)}+\sum_{n=1}^{\infty}\left(\frac{2}{n}\right)\left(\frac{q^{n}}{1+q^{2n}}\right)\sin\left(\frac{n\pi u}{{\cal K}_{1}(p)}\right),
\end{equation}
where $q$ is known as the \textit{nome} and is given by $q=\exp[-\pi {\cal K}_{1}(\sqrt{1-p^{2}})/{\cal K}_{1}(p)]$.
\section{Perturbation Results}
\label{sec:pertapp}
The perturbation results up to second order in $\delta K_{1}$, $\delta K_{2}$ and $\alpha$ for the tilt angle $\phi$ are presented below.
\begin{eqnarray}
\phi_{1\alpha}(\zt)&=&-\frac{{\cal E}^{3}}{2(1+{\cal E}^{2})^{\frac{3}{2}}\chi}\frac{(2\chi+\sqrt{1+{\cal E}^{2}})}{(4\chi+\sqrt{1+{\cal E}^{2}})}\nonumber\\&&-\frac{{\cal E}^{3}}{2(1+{\cal E}^{2})\chi}\frac{\cos(4\pi \zt)}{(4\chi+\sqrt{1+{\cal E}^{2}})}
\end{eqnarray}
\begin{eqnarray}
\phi_{2\alpha}(\zt)&=&\frac{{\cal E}^{3}}{2(1+{\cal E}^{2})\chi}\frac{(1+2\cos(4\pi \zt))}{(\sqrt{1+{\cal E}^{2}}+4\chi)}
\end{eqnarray}
\begin{eqnarray}
\phi_{11}(\zt)&=&\frac{\chi{\cal E}^{3}}{2(1+{\cal E}^{2})^{\frac{5}{2}}}\frac{(1+{\cal E}^{2}+6\sqrt{1+{\cal E}^{2}}\chi+8\chi^{2})}{(1+{\cal E}^{2}+8\chi(\sqrt{1+{\cal E}^{2}}+2\chi))}\nonumber\\&&+\frac{\chi{\cal E}^{3}}{2(1+{\cal E}^{2})^{2}}\frac{(\sqrt{1+{\cal E}^{2}}+4\chi)\cos(4\pi \zt)}{(1+{\cal E}^{2}+8\chi(\sqrt{1+{\cal E}^{2}}+2\chi))}\nonumber\\
\end{eqnarray}
\begin{eqnarray}
\phi_{12}(\zt)&=&-\frac{{\cal E}^{3}\chi}{2(1+{\cal E}^{2})^{2}}\frac{(\sqrt{1+{\cal E}^{2}}+12\chi)(1+2\cos(4\pi \zt))}{(1+{\cal E}^{2}+16\chi(\sqrt{1+{\cal E}^{2}}+3\chi))}\nonumber\\
\end{eqnarray}
\begin{eqnarray}
\phi_{22}(\zt)&=&{\cal E}\frac{(1+\chi^{2}+{\cal E}^{2}(1-\chi^{2})+\chi({\cal E}^{2}-2)\sqrt{1+{\cal E}^{2}})}{(1+{\cal E}^{2})^{3}}\nonumber\\
\end{eqnarray}


\begin{thebibliography}{10}%
\makeatletter
\providecommand \@ifxundefined [1]{%
 \ifx #1\undefined \expandafter \@firstoftwo
 \else \expandafter \@secondoftwo
\fi
}%
\providecommand \@ifnum [1]{%
 \ifnum #1\expandafter \@firstoftwo
 \else \expandafter \@secondoftwo
\fi
}%
\providecommand \enquote [1]{``#1''}%
\providecommand \bibnamefont  [1]{#1}%
\providecommand \bibfnamefont [1]{#1}%
\providecommand \citenamefont [1]{#1}%
\providecommand\href[0]{\@sanitize\@href}%
\providecommand\@href[1]{\endgroup\@@startlink{#1}\endgroup\@@href}%
\providecommand\@@href[1]{#1\@@endlink}%
\providecommand \@sanitize [0]{\begingroup\catcode`\&12\catcode`\#12\relax}%
\@ifxundefined \pdfoutput {\@firstoftwo}{%
 \@ifnum{\z@=\pdfoutput}{\@firstoftwo}{\@secondoftwo}%
}{%
 \providecommand\@@startlink[1]{\leavevmode}%
 \providecommand\@@endlink[0]{}%
}{%
 \providecommand\@@startlink[1]{%
  \leavevmode
  \pdfstartlink
   attr{/Border[0 0 1 ]/H/I/C[0 1 1]}%
   user{/Subtype/Link/A<</Type/Action/S/URI/URI(#1)>>}%
  \relax
 }%
 \providecommand\@@endlink[0]{\pdfendlink}%
}%
\providecommand \url  [0]{\begingroup\@sanitize \@url }%
\providecommand \@url [1]{\endgroup\@href {#1}{\urlprefix}}%
\providecommand \urlprefix [0]{URL }%
\providecommand \Eprint[0]{\href }%
\@ifxundefined \urlstyle {%
  \providecommand \doi [1]{doi:\discretionary{}{}{}#1}%
}{%
  \providecommand \doi [0]{doi:\discretionary{}{}{}\begingroup
  \urlstyle{rm}\Url }%
}%
\providecommand \doibase [0]{http://dx.doi.org/}%
\providecommand \Doi[1]{\href{\doibase#1}}%
\providecommand \bibAnnote [3]{%
  \BibitemShut{#1}%
  \begin{quotation}\noindent
    \textsc{Key:}\ #2\\\textsc{Annotation:}\ #3%
  \end{quotation}%
}%
\providecommand \bibAnnoteFile [2]{%
  \IfFileExists{#2}{\bibAnnote {#1} {#2} {\input{#2}}}{}%
}%
\providecommand \typeout [0]{\immediate \write \m@ne }%
\providecommand \selectlanguage [0]{\@gobble}%
\providecommand \bibinfo [0]{\@secondoftwo}%
\providecommand \bibfield [0]{\@secondoftwo}%
\providecommand \translation [1]{[#1]}%
\providecommand \BibitemOpen[0]{}%
\providecommand \bibitemStop [0]{}%
\providecommand \bibitemNoStop [0]{.\EOS\space}%
\providecommand \EOS [0]{\spacefactor3000\relax}%
\providecommand \BibitemShut [1]{\csname bibitem#1\endcsname}%
\bibitem{Meyer:69}%
  \BibitemOpen
  \bibfield{author}{%
  \bibinfo {author} {\bibfnamefont{R.~B.}\ \bibnamefont{Meyer}},\ }%
  \bibfield{journal}{%
  \bibinfo {journal} {Phys. Rev. Lett.}\ }%
  \textbf{\bibinfo {volume} {22}},\ \bibinfo {pages} {918} (\bibinfo {year}
  {1969})%
  \bibAnnoteFile{NoStop}{Meyer:69}%
\bibitem{Rudquist:97}%
  \BibitemOpen
  \bibfield{author}{%
  \bibinfo {author} {\bibfnamefont{P.}~\bibnamefont{Rudquist}}, \bibinfo
  {author} {\bibfnamefont{T.}~\bibnamefont{Carlsson}}, \bibinfo {author}
  {\bibfnamefont{L.}~\bibnamefont{Komitov}},\ and\ \bibinfo {author}
  {\bibfnamefont{S.~T.}\ \bibnamefont{Lagerwall}},\ }%
  \bibfield{journal}{%
  \bibinfo {journal} {Liquid Crystals}\ }%
  \textbf{\bibinfo {volume} {22}},\ \bibinfo {pages} {445} (\bibinfo {year}
  {1997})%
  \bibAnnoteFile{NoStop}{Rudquist:97}%
\bibitem{Patel:87}%
  \BibitemOpen
  \bibfield{author}{%
  \bibinfo {author} {\bibfnamefont{J.~S.}\ \bibnamefont{Patel}}\ and\ \bibinfo
  {author} {\bibfnamefont{R.~B.}\ \bibnamefont{Meyer}},\ }%
  \bibfield{journal}{%
  \bibinfo {journal} {Phys. Rev. Lett.}\ }%
  \textbf{\bibinfo {volume} {58}},\ \bibinfo {pages} {1538} (\bibinfo {year}
  {1987})%
  \bibAnnoteFile{NoStop}{Patel:87}%
\bibitem{Rudquist:98a}%
  \BibitemOpen
  \bibfield{author}{%
  \bibinfo {author} {\bibfnamefont{P.}~\bibnamefont{Rudquist}}, \bibinfo
  {author} {\bibfnamefont{L.}~\bibnamefont{Komitov}},\ and\ \bibinfo {author}
  {\bibfnamefont{S.~T.}\ \bibnamefont{Lagerwall}},\ }%
  \bibfield{journal}{%
  \bibinfo {journal} {Liquid Crystals}\ }%
  \textbf{\bibinfo {volume} {24}},\ \bibinfo {pages} {329} (\bibinfo {year}
  {1998})%
  \bibAnnoteFile{NoStop}{Rudquist:98a}%
\bibitem{Coles:06}%
  \BibitemOpen
  \bibfield{author}{%
  \bibinfo {author} {\bibfnamefont{B.~J.}\ \bibnamefont{Broughton}}, \bibinfo
  {author} {\bibfnamefont{M.~J.}\ \bibnamefont{Clarke}}, \bibinfo {author}
  {\bibfnamefont{S.~M.}\ \bibnamefont{Morris}}, \bibinfo {author}
  {\bibfnamefont{A.~E.}\ \bibnamefont{Blatch}},\ and\ \bibinfo {author}
  {\bibfnamefont{H.~J.}\ \bibnamefont{Coles}},\ }%
  \bibfield{journal}{%
  \bibinfo {journal} {J. App. Phys.}\ }%
  \textbf{\bibinfo {volume} {99}},\ \bibinfo {pages} {023511} (\bibinfo {year}
  {2006})%
  \bibAnnoteFile{NoStop}{Coles:06}%
\bibitem{Lee:90}%
  \BibitemOpen
  \bibfield{author}{%
  \bibinfo {author} {\bibfnamefont{S.~D.}\ \bibnamefont{Lee}}, \bibinfo
  {author} {\bibfnamefont{J.~S.}\ \bibnamefont{Patel}},\ and\ \bibinfo {author}
  {\bibfnamefont{R.~B.}\ \bibnamefont{Meyer}},\ }%
  \bibfield{journal}{%
  \bibinfo {journal} {J. App. Phys.}\ }%
  \textbf{\bibinfo {volume} {67}},\ \bibinfo {pages} {1293} (\bibinfo {year}
  {1990})%
  \bibAnnoteFile{NoStop}{Lee:90}%
\bibitem{Rudquist:94}%
  \BibitemOpen
  \bibfield{author}{%
  \bibinfo {author} {\bibfnamefont{P.}~\bibnamefont{Rudquist}}, \bibinfo
  {author} {\bibfnamefont{L.}~\bibnamefont{Komitov}},\ and\ \bibinfo {author}
  {\bibfnamefont{S.~T.}\ \bibnamefont{Lagerwall}},\ }%
  \bibfield{journal}{%
  \bibinfo {journal} {Phys. Rev. E}\ }%
  \textbf{\bibinfo {volume} {50}},\ \bibinfo {pages} {4735} (\bibinfo {year}
  {1994})%
  \bibAnnoteFile{NoStop}{Rudquist:94}%
\bibitem{petrov}%
  \BibitemOpen
  \bibfield{author}{%
  \bibinfo {author} {\bibfnamefont{A.~G.}\ \bibnamefont{Petrov}},\ }%
  \enquote{\bibinfo {title} {Measurements and interpretation of
  flexoelectricity},}\ in\ \emph{\bibinfo {booktitle} {Physical Properties of
  Liquid Crystals}},\ \bibinfo {editor} {edited by\ \bibinfo {editor}
  {\bibfnamefont{A.~F.}\ \bibnamefont{D.~A.~Dunmur}}\ and\ \bibinfo {editor}
  {\bibfnamefont{G.~R.}\ \bibnamefont{Luckhurst}}}\ (\bibinfo {publisher}
  {IEE},\ \bibinfo {address} {London},\ \bibinfo {year} {2001})\ pp.\ \bibinfo
  {pages} {251--264}%
  \bibAnnoteFile{NoStop}{petrov}%
\bibitem{deGennesprostbook:93}%
  \BibitemOpen
  \bibfield{author}{%
  \bibinfo {author} {\bibfnamefont{P.~G.}\ \bibnamefont{de~Gennes}}\ and\
  \bibinfo {author} {\bibfnamefont{J.}~\bibnamefont{Prost}},\ }%
  \emph{\bibinfo {title} {The Physics of Liquid Crystals}}\ (\bibinfo
  {publisher} {Oxford University Press},\ \bibinfo {address} {Oxford},\
  \bibinfo {year} {1993})%
  \bibAnnoteFile{NoStop}{deGennesprostbook:93}%
\bibitem{Brimicombe:07}%
  \BibitemOpen
  \bibfield{author}{%
  \bibinfo {author} {\bibfnamefont{P.~D.}\ \bibnamefont{Brimicombe}}, \bibinfo
  {author} {\bibfnamefont{C.}~\bibnamefont{Kischka}}, \bibinfo {author}
  {\bibfnamefont{S.~J.}\ \bibnamefont{Elston}},\ and\ \bibinfo {author}
  {\bibfnamefont{E.~P.}\ \bibnamefont{Raynes}},\ }%
  \bibfield{journal}{%
  \bibinfo {journal} {J. Appl. Phys.}\ }%
  \textbf{\bibinfo {volume} {101}},\ \bibinfo {pages} {043108} (\bibinfo {year}
  {2007})%
  \bibAnnoteFile{NoStop}{Brimicombe:07}%
\bibitem{Patrick:09}%
  \BibitemOpen
  \bibfield{author}{%
  \bibinfo {author} {\bibfnamefont{P.~S.}\ \bibnamefont{Salter}}, \bibinfo
  {author} {\bibfnamefont{C.}~\bibnamefont{Kischka}}, \bibinfo {author}
  {\bibfnamefont{S.~J.}\ \bibnamefont{Elston}},\ and\ \bibinfo {author}
  {\bibfnamefont{E.~P.}\ \bibnamefont{Raynes}},\ }%
  \bibfield{journal}{%
  \bibinfo {journal} {Liquid Crystals}\ }%
  \textbf{\bibinfo {volume} {36}},\ \bibinfo {pages} {1355} (\bibinfo {year}
  {2009})%
  \bibAnnoteFile{NoStop}{Patrick:09}%
\bibitem{Davidson:02}%
  \BibitemOpen
  \bibfield{author}{%
  \bibinfo {author} {\bibfnamefont{A.~J.}\ \bibnamefont{Davidson}}\ and\
  \bibinfo {author} {\bibfnamefont{N.~J.}\ \bibnamefont{Mottram}},\ }%
  \bibfield{journal}{%
  \bibinfo {journal} {Phys. Rev. E}\ }%
  \textbf{\bibinfo {volume} {65}},\ \bibinfo {pages} {051710} (\bibinfo {year}
  {2002})%
  \bibAnnoteFile{NoStop}{Davidson:02}%
\bibitem{Coles:01}%
  \BibitemOpen
  \bibfield{author}{%
  \bibinfo {author} {\bibfnamefont{H.~J.}\ \bibnamefont{Coles}}, \bibinfo
  {author} {\bibfnamefont{B.}~\bibnamefont{Musgrave}}, \bibinfo {author}
  {\bibfnamefont{M.~J.}\ \bibnamefont{Coles}},\ and\ \bibinfo {author}
  {\bibfnamefont{J.}~\bibnamefont{Willmott}},\ }%
  \bibfield{journal}{%
  \bibinfo {journal} {J. Mater. Chem.}\ }%
  \textbf{\bibinfo {volume} {11}},\ \bibinfo {pages} {2709} (\bibinfo {year}
  {2001})%
  \bibAnnoteFile{NoStop}{Coles:01}%
\bibitem{Cole06a}%
  \BibitemOpen
  \bibfield{author}{%
  \bibinfo {author} {\bibfnamefont{H.~J.}\ \bibnamefont{Coles}}, \bibinfo
  {author} {\bibfnamefont{M.~J.}\ \bibnamefont{Clarke}}, \bibinfo {author}
  {\bibfnamefont{S.~M.}\ \bibnamefont{Morris}}, \bibinfo {author}
  {\bibfnamefont{B.~J.}\ \bibnamefont{Broughton}},\ and\ \bibinfo {author}
  {\bibfnamefont{A.~E.}\ \bibnamefont{Blatch}},\ }%
  \bibfield{journal}{%
  \bibinfo {journal} {J. Appl. Phys.}\ }%
  \textbf{\bibinfo {volume} {99}},\ \bibinfo {pages} {034104} (\bibinfo {year}
  {2006})%
  \bibAnnoteFile{NoStop}{Cole06a}%
\bibitem{castles:09}%
  \BibitemOpen
  \bibfield{author}{%
  \bibinfo {author} {\bibfnamefont{F.}~\bibnamefont{Castles}}, \bibinfo
  {author} {\bibfnamefont{S.~M.}\ \bibnamefont{Morris}},\ and\ \bibinfo
  {author} {\bibfnamefont{H.~J.}\ \bibnamefont{Coles}},\ }%
  \bibfield{journal}{%
  \bibinfo {journal} {Phys. Rev. E}\ }%
  \textbf{\bibinfo {volume} {80}},\ \bibinfo {pages} {031709} (\bibinfo {year}
  {2009})%
  \bibAnnoteFile{NoStop}{castles:09}%
\end{thebibliography}
\end{document}